\documentclass[11pt,a4paper,final]{article}
\usepackage{amsmath, amssymb, amsthm}
\usepackage{graphicx}
\usepackage{lscape}
\usepackage{subfigure}

\title{Global QCD fit from $Q^2=0$ to $Q^2=30000$ GeV$^2$ with Regge-compatible initial condition}
\author{G. Soyez}

\newcommand{\cA}{{\cal A}}
\newcommand{\cB}{{\cal B}}
\newcommand{\cC}{{\cal C}}
\newcommand{\cD}{{\cal D}}
\newcommand{\cb}{b}
\newcommand{\cg}{\gamma}

\begin{document}

\maketitle

\begin{abstract}
  In this paper I show that it is possible to use Regge theory to constrain the initial parton distribution functions of a global DGLAP fit. In this approach, both quarks and gluons have the same high-energy behaviour which may also be used to describe soft interactions. More precisely, I show that, if we parametrise the parton distributions with a triple-pole pomeron, {\em i.e.} like $\log^2(1/x)$ at small $x$, at $Q^2=Q_0^2$ and evolve these distribution with the DGLAP equation, we can reproduce $F_2^p$, $F_2^d$, $F_2^n/F_2^p$, $F_2^{\nu N}$ and $xF_3^{\nu N}$ for $W^2\ge 12.5$ GeV$^2$. In this case, we obtain a new leading-order global QCD fit with a Regge-compatible initial condition.

I shall also show that it is possible to use Regge theory to extend the parton distribution functions to small $Q^2$. This leads to a description of the structure functions over the whole $Q^2$ range based on Regge theory at low $Q^2$ and on QCD at large $Q^2$. 

Finally, I shall argue that, at large $Q^2$, the parton distribution functions obtained from DGLAP evolution and containing an essential singularity at $j=1$ can be approximated by a triple-pole pomeron behaviour.
\end{abstract}

\section{Introduction}

About thirty years ago, Dokshitzer, Gribov, Lipatov, Altarelli and Parisi have shown \cite{DGLAP} that quantum chromodynamics predicts a breakdown of Bjorken scaling in Deep Inelastic Scattering (DIS). Once the parton distribution functions are fixed at one initial scale $Q^2=Q_0^2$, the DGLAP equation gives their evolution to larger values of $Q^2$. Although the initial equation has only included QCD contributions at leading order (LO) in $\alpha_s$, the next-to-leading order (NLO) corrections are now known as well as the NNLO corrections.
There exists a rather large number of global fits (e.g. \cite{Martin:2001es,Gluck:1998xa,Pumplin:2002vw,Cooper-Sarkar:2002yx,Alekhin:2000ch}) using the DGLAP equation to reproduce the DIS data. The basic idea is to fix the initial parton distributions, not predicted by perturbative QCD (pQCD), and to evolve it in order to reproduce the experimental measurements as well as possible. The success of this type of analysis is often considered as one of the most important prediction of pQCD. However, it appears that this approach presents some problems. Firstly, even if the strong rise of $F_2$ observed by HERA at small $x$ is well reproduced by the DGLAP evolution, this may seem surprising since the evolution generates an unphysical essential singularity which should be replaced by the BFKL small-$x$ behaviour \cite{bfkl} which is not observed in the data and is unstable against NLO corrections \cite{Fadin:1989kf}. In addition, since the initial parton distributions are not predicted by pQCD, we have to parametrise them. This introduces a large number of free parameters in the models.

In this paper, we shall study the possibility to use Regge theory \cite{books,Regge:mz,Regge:1960zc} to constrain the initial parton distributions used in QCD global fits. To motivate this approach, one can, for example, consider the MRST2002 initial conditions: at small $x$, we have
\begin{eqnarray*}
xq(x, Q_0^2) & = & A x^{-0.12},\\
xg(x, Q_0^2) & = & B x^{-0.27} + C x^{0.00}.
\end{eqnarray*}
These singularities in $x$ do not correspond to any singularity present in hadronic cross sections \cite{RPP,Donnachie:2001xx,Cudell:2001ii,Desgrolard:2001bu} and, conversely, cross-section singularities are not present in parton distributions. There should therefore exist a mechanism explaining how the residues of these singularities in partonic distributions vanish when $Q^2$ goes to zero, and how the residues of the singularities observed in the total cross sections vanish for nonzero $Q^2$. Such a mechanism is unknown and seems forbidden in Regge theory, hence a description of both total cross sections and partonic distributions with the same singularity structure seems necessary.

In a previous work \cite{Soyez:2002nm}, we have shown that, if one considers only the small-$x$ and large-$Q^2$ domain, on can use a triple-pole pomeron (squared logarithm of $x$) to reproduce the low-$Q^2$ data and evolve it using the DGLAP equation to obtain the high-$Q^2$ measurements. More precisely, we have used parametrisations of the form
\[
A \log^2(1/x) + B \log(1/x) + C + D \left(\frac{1}{x}\right)^{-\eta}
\]
for initial quark and gluon distributions. Since this Regge-constrained parametrisation does not extend to $x=1$, we have used the GRV parton distributions at large $x$ ($x \ge x_{\text{Regge}} \approx 0.15$). As a consequence, only two quark distributions were needed (a flavour singlet coupled to gluons and a flavour non-singlet) and only the proton structure function was fitted. 

I£n this paper, we shall extend the triple-pole parametrisation up to $x=1$ in order to obtain a global QCD fit compatible with Regge theory at small $Q^2$. At first sight, if we want to replace the GRV parton distributions at large $x$, we can, for example, introduce powers of $(1-x)$. However, if we do so, we must not only concentrate on the Regge domain but also on the large-$x$ experimental measurements, including other experiments like $\gamma^* d$ scattering and the neutrino data. We shall fit all structure functions over the whole $x$ range at all scales greater than $Q_0^2$. This will allow us to extract the parton distribution functions, which will be parametrised in such a way that they agree with Regge theory at small values of $x$.

In this study, we shall also consider the extension of the Regge initial parametrisation to low-$Q^2$ values. Using standard techniques, we shall show that it is possible to continue the initial parton distributions to small $Q^2$.

Finally, we shall show that, at large $Q^2$, the parton distributions obtained from DGLAP evolution, containing an essential singularity, can be approximated by a squared logarithm of $1/x$ within estimated DGLAP uncertainties. This confirms the results obtained in \cite{Soyez:2003sr} where we have shown that the residues of the triple-pole pomeron can be extracted from DGLAP evolution.

We should also point out that a similar type of approach has already been used by Donnachie and Landshoff \cite{Donnachie:2001zt}, and by Csernai, Jenkovszky, Kontros, Lengyel Magas and Paccanoni \cite{Csernai:2001fk}. However, our approach is different from their work in a number of points. Firstly, we keep the full DGLAP evolution equation, without taking only the residue at the leading singularity as done by Donnachie and Landshoff. In addition, the work by Csernai {\em et al.} only reproduces $F_2^p$ which means that they only need two quark distributions: a flavour singlet and a flavour non-singlet, only the flavour-singlet distribution being parametrised using Regge theory. Ther present work extends these ideas and those of \cite{Soyez:2002nm,Soyez:2003sr} to provide a standard set\footnote{A C code for this standard set is available at http://lepton.theo.phys.ulg.ac.be/$\tilde{\phantom{i}}$soyez.} of structure functions, which reproduces the data for all values of $x$ and $Q^2$.

\section{Fitted and evolved quantities}

If we want to extend the parametrisation introduced in \cite{Soyez:2002nm} up to $x=1$, we cannot only restrict ourselves to $F_2^p$. In order to have a good determination of the valence quarks and of the sea asymmetry, we also need to include other structure functions, measured in the large-$x$ region. In this global fit, we thus include the following quantities:
\begin{itemize}
\item The proton structure function $F_2^p$ \cite{Abt:1993cb,Ahmed:1995fd,Aid:1996au,Adloff:1997mf,Adloff:1999ah,Adloff:2000qj,Adloff:2000qk,Derrick:1993ft,Derrick:1994sz,Derrick:1995ef,Derrick:1996hn,Breitweg:1997hz,Breitweg:1998dz,Breitweg:1999ad,Chekanov:2001qu,Benvenuti:1989rh,Adams:1996gu,Arneodo:1996qe,Whitlow:1991uw}: this is by far the most important type of experimental data. Moreover, it is nearly the only one to contribute to the fit in the small-$x$ or in the high-$Q^2$ region.
\item The deuteron structure function $F_2^d$ \cite{Adams:1996gu,Arneodo:1996qe,Whitlow:1991uw,Benvenuti:1989gs}: as we shall see, these data allow the determination of the sea asymmetry. Many points are available in the large- and middle-$x$ regions, where the sea asymmetry is expected to be large.
\item The neutrino structure functions $F_2^{\nu N}$ and $F_3^{\nu N}$ \cite{Oltman:pq,Seligman:mc,Fleming:2000bg}: these data, in which most of the points are at large values of $x$, are important to fix the strange quark and the valence quark distributions. Note that the data considered here are averaged over neutrinos and anti-neutrinos.
\item The $F_2^n/F_2^p$ measurements \cite{Amaudruz:1991nw}: these data constrain the valence quark distributions and the sea asymmetry.
\end{itemize}

Once we know which experiments are fitted, we must find which quantities need to be evolved. Since the $Q^2$ range under consideration in global fits extends up to 30000 GeV$^2$, we need to consider 5 quark flavours: $u$, $d$, $s$, $c$ and $b$. In order to use DGLAP evolution, it is easier to perform linear combinations of the quark distributions. In our case, we shall use 6 flavour-non-singlet distributions
\begin{eqnarray}\label{eq:nsdists}
xu_V   & = & x(u - \bar{u}),\nonumber\\
xd_V   & = & x(d - \bar{d}),\nonumber\\
T_3    & = & x(u^+ - d^+),\nonumber\\[-3mm]
& & \\[-3mm]
T_8    & = & x(u^+ + d^+ - 2 s^+),\nonumber\\
T_{15} & = & x(u^+ + d^+ + s^+ - 3 c^+),\nonumber\\
T_{24} & = & x(u^+ + d^+ + s^+ + c^+ - 4b^+),\nonumber
\end{eqnarray}
where $q^+ = q+\bar{q}$. Note that since the proton does not contain constituent strange, charm or bottom valence quarks, we have $s=\bar{s}$, $c=\bar{c}$ and $b=\bar{b}$. At leading order, the $Q^2$ evolution of each of these distributions is given by the DGLAP equation with the splitting $xP_{qq}(x)$. In addition to the non-singlet distributions, we have the singlet quark distribution
\[
\Sigma = x(u^+ + d^+ + s^+ + c^+ + b^+)
\]
which evolves coupled to the gluon distribution $G=xg$, with the full splitting matrix
\[
\begin{pmatrix}
xP_{qq}(x) & 2n_fxP_{qg}(x)\\
xP_{gq}(x) & xP_{gg}(x)
\end{pmatrix}.
\]
We shall assume that for $Q^2\le 4m_q^2$, the quark $q$ does not enter into the evolution equations.

If we invert the relations \eqref{eq:nsdists} and express the quark densities $q^+$ in terms of the evolved quantities, we obtain
\begin{eqnarray*}
xu^+ & = & \frac{1}{60}(12\Sigma + 3T_{24} + 5T_{15} + 10T_8 + 30T_3),\\
xd^+ & = & \frac{1}{60}(12\Sigma + 3T_{24} + 5T_{15} + 10T_8 - 30T_3),\\
xs^+ & = & \frac{1}{60}(12\Sigma + 3T_{24} + 5T_{15} - 20T_8),\\
xc^+ & = & \frac{1}{20}( 4\Sigma +  T_{24} - 5T_{15}),\\
xb^+ & = & \frac{1}{ 5}(  \Sigma -  T_{24}).
\end{eqnarray*}

Now, we can of course write the structure functions considered here in terms of the parton distributions or in terms of the flavour-singlet and flavour-non-singlet distributions\footnote{At leading order, the quark coefficient functions are proportional to $\delta(1-x)$ and the gluon coefficient function vanishes.}. If, for the sake of clarity, we include other quantities like the neutron structure function, this gives
\begin{eqnarray*}
F_2^p & = & \frac{4x}{9}(u^++c^+)+\frac{x}{9}(d^++s^++b^+) \\
      & = & \frac{1}{90}(22\Sigma+3T_{24}-5T_{15}+5T_8+15T_3),\\
F_2^n & = & \frac{4x}{9}(d^++c^+)+\frac{x}{9}(u^++s^++b^+) \\
      & = & \frac{1}{90}(22\Sigma+3T_{24}-5T_{15}+5T_8-15T_3),\\
F_2^d & = & \frac{F_2^p+F_2^n}{2}\\
      & = & \frac{5x}{18}(u^++d^+)+\frac{4x}{9}c^++\frac{x}{9}(s^++b^+)\\
      & = & \frac{1}{90}(22\Sigma+3T_{24}-5T_{15}+5T_8),\\
\end{eqnarray*}
and for the neutrino structure functions
\begin{eqnarray*}
F_2^{\nu p} & = & 2x(d+s+b+\bar{u}+\bar{c}),\\
F_2^{\nu n} & = & 2x(u+s+b+\bar{d}+\bar{c}),\\
F_2^{\bar{\nu} p} & = & 2x(u+c+\bar{d}+\bar{s}+\bar{b}),\\
F_2^{\bar{\nu} n} & = & 2x(d+c+\bar{u}+\bar{s}+\bar{b}),\\
xF_3^{\nu p} & = & 2x(d+s+b-\bar{u}-\bar{c}),\\
xF_3^{\nu n} & = & 2x(u+s+b-\bar{d}-\bar{c}),\\
xF_3^{\bar{\nu} p} & = & 2x(u+c-\bar{d}-\bar{s}-\bar{b}),\\
xF_3^{\bar{\nu} n} & = & 2x(d+c-\bar{u}-\bar{s}-\bar{b}),\\
\end{eqnarray*}
If we average over proton and neutron targets, we obtain the neutrino-nucleon structure functions\footnote{Neutrino experiments are often performed with heavy nuclei which means that the averaged structure function is measured.}
\begin{eqnarray*}
F_2^{\nu N} = F_2^{\bar{\nu} N} & = & x(u^++d^++s^++c^++b^+),\\
xF_3^{\nu N} & = & x(u_V+d_V+s^+-c^++b^+),\\
xF_3^{\bar{\nu} N} & = & x(u_V+d_V-s^++c^+-b^+).\\
\end{eqnarray*}
We may finally average over neutrinos and anti-neutrinos, which leads to
\begin{eqnarray*}
F_2^{\stackrel{(-)}\nu N} & = & x(u^++d^++s^++c^+),\\
                          & = & \Sigma,\\
xF_3^{\stackrel{(-)}\nu N} & = & x(u_V+d_V).
\end{eqnarray*}

\section{Initial parametrisation}

If we want to perform a DGLAP evolution, we need to fix the parton distribution functions at an initial scale $Q_0^2$. Following the same ideas as in \cite{Soyez:2002nm}, we shall parametrise each quark distribution as the sum of a triple-pole pomeron term and an $a_2/f$-reggeon term. In addition, each distribution will be multiplied by a power of $(1-x)$, to ensure that the parametrisation extended to $x=1$ goes to 0 when $x\to 1$. This leads to the following parametrisation
\[
xq(x,Q_0^2) = \left[A_q \log^2(1/x) + B_q \log(1/x) + C_q + D_q x^\eta\right] (1-x)^{b_q},
\]
with\footnote{The sea distribution $q_s$ is simply $\frac{1}{2}q^+$.} $q=u_V$, $d_V$, $u_s$, $d_s$, $s_s$, $c_s$ and $g$. Fortunately, we can restrict many of the 35 parameters introduced here:
\begin{itemize}
\item First of all, the charm (bottom) distribution will be set to zero for $Q^2\le~4m_c^2$ ($Q^2\le~4m_b^2$). We shall therefore take $Q_0^2\le~4m_c^2$ so that we can set $c(x, Q_0^2)=0$ and $b(x, Q_0^2)=0$. In other words, we have $T_{15}(x,Q^2)=\Sigma(x,Q^2)$ for $Q^2\le~4m_c^2$ and $T_{24}(x,Q^2)=\Sigma(x,Q^2)$ for $Q^2\le~4m_b^2$.
\item The pomeron does not distinguish between quarks and anti-quarks. This means that the valence distributions $u_V$ and $d_V$ do not contain a pomeron term.
\item The pomeron, having vacuum quantum numbers, is insensitive to quark flavour. Thus, the only parameter through which the quark flavour may influence the pomeron is its mass. In other words, the couplings $A_q$, $B_q$ and $C_q$ are functions of $Q^2$ and $m_q^2$ only. Consequently, the pomeron contributions to the $u_s$ and $d_s$ densities are the same. Assuming that the strange mass is very small compared to the virtualities $Q^2$ under consideration, we shall also take the same pomeron contribution\footnote{If we insert an overall factor in $s_s$, the fit naturally sets it to 1.} in $s_s$.
\begin{eqnarray*}
A_u = A_d =  A_s = A,\\
B_u = B_d =  B_s = B,\\
C_u = C_d =  C_s = C.
\end{eqnarray*}
\item We shall assume that the reggeon, being mainly constituted of quarks, does not contribute to the gluon distribution. The parameter $D_g$ will thus\footnote{If we do not impose $D_g=0$, the parameter stays small in the fit.} be set to 0.
\item We know from \cite{Lopez:1979bb} that, at large $x$, the following behaviour is stable with respect to DGLAP evolution
\begin{eqnarray*}
\Sigma & \sim & (1-x)^b,\\
G      & \sim & \frac{(1-x)^{b+1}}{\log\left(\frac{1}{1-x}\right)}.
\end{eqnarray*}
The denominator $\log(1-x)$ in the gluon distribution does not have a good behaviour at small $x$ so we have not included it\footnote{One solution is to multiply the gluon distribution by an overall factor $x$. This makes no change at large $x$ and ensures a good behaviour at small $x$ because, when $x\to 0$,
\[
\frac{x}{\log\left(\frac{1}{1-x}\right)} \to 1.
\]
Numerically, including this factor in the gluon distribution only makes a small correction.}. Nevertheless, we shall impose
\begin{eqnarray*}
& b_u = b_d = b_s = b,& \\
& b_g = b+1.&
\end{eqnarray*}
\item If we look at the large-$x$ data, we can see that if we only use $Dx^\eta(1-x)^b$ for the valence quarks, the resulting distribution is too wide, or has a peak at too small a value of $x$. In order to solve that problem, we have multiplied the valence-quark distributions by a factor $(1+\gamma_q x)$.
\item Finally, we still need to impose sumrules. Quark-number conservation can be used to fix the valence-quark normalisation factors. If we write
\[
A_{u_V} = \frac{2}{N_u} \qquad\text{and}\qquad A_{d_V} = \frac{1}{N_d},
\]
we find
\begin{equation}\label{eq:qnumc}
N_q = \frac{\Gamma(b_q+1)\Gamma(\eta)}{\Gamma(\eta+b_q+1)}\left(1+\frac{\gamma_q\eta}{\eta+b_q+1}\right).
\end{equation}
The momentum sumrule is used to fix the constant term $C_g$ in the gluon distribution. Although all the functions involved are analytically integrable, the resulting expression for $C_g$ is quite complicated and we give it in appendix.
\end{itemize}

Taking all these considerations into account, we obtain the following parametrisation for the initial distributions
\begin{eqnarray}
xu_V & = & \frac{2}{N_u^*} x^\eta(1+\gamma_u x)(1-x)^{b_u},\nonumber\\
xd_V & = & \frac{1}{N_d^*} x^\eta(1+\gamma_d x)(1-x)^{b_d},\nonumber\\
xu_s & = & \left[A\log^2(1/x)+B\log(1/x)+C+D_u x^\eta \right](1-x)^b,\nonumber\\
xd_s & = & \left[A\log^2(1/x)+B\log(1/x)+C+D_d x^\eta \right](1-x)^b,\nonumber\\[-3mm]
&&\label{eq:initdistrib}\\[-3mm]
xs_s & = & \left[A\log^2(1/x)+B\log(1/x)+C+D_s x^\eta \right](1-x)^b,\nonumber\\
xc_s & = & 0,\nonumber\\
xb_s & = & 0,\nonumber\\
xg   & = & \left[A_g\log^2(1/x)+B_g\log(1/x)+C_g^*\right](1-x)^{b+1},\nonumber
\end{eqnarray}
where the parameters marked with an asterisk are constrained by sumrules.

\section{Fitted experiments}

As said previously, we have fitted $F_2^p$, $F_2^d$, $F_2^{\nu N}$, $xF_3^{\nu N}$ and $F_2^n/F_2^p$. We shall now detail which experiments are included in the fit for all these quantities.

For the proton structure function, we have fitted the experiments from\footnote{The dataset is coming from the DURHAM database (http://durpdg.dur.ac.uk) to which we have added the 2000 and 2001 data from HERA \cite{Adloff:2000qj,Adloff:2000qk,Chekanov:2001qu} as well as the reanalysed CCFR 2001 data \cite{Fleming:2000bg}.} H1 \cite{Abt:1993cb,Ahmed:1995fd,Aid:1996au,Adloff:1997mf,Adloff:1999ah,Adloff:2000qj,Adloff:2000qk}, ZEUS \cite{Derrick:1993ft,Derrick:1994sz,Derrick:1995ef,Derrick:1996hn,Breitweg:1997hz,Breitweg:1998dz,Breitweg:1999ad,Chekanov:2001qu}, BCDMS \cite{Benvenuti:1989rh}, E665 \cite{Adams:1996gu}, NMC \cite{Arneodo:1996qe} and SLAC \cite{Whitlow:1991uw}. For the deuteron structure function measurements, we have included data from BCDMS \cite{Benvenuti:1989gs} E665 \cite{Adams:1996gu} and NMC \cite{Arneodo:1996qe}. We have also taken into account the measurements of $F_2^n/F_2^p$ from NMC \cite{Amaudruz:1991nw}. Finally, the neutrino data used here come from CCFR \cite{Oltman:pq,Seligman:mc,Fleming:2000bg}.

Among all these experimental papers, some give, besides the statistical and the systematic errors, an additional normalisation uncertainty. For each of these subsets of the data, we have allowed an overall normalisation factor. Let $R_i$ be the normalisation uncertainty for the subset $i$, and $\rho_i$ the effective normalisation factor. We may easily minimise the $\chi^2$ with respect to this parameter by requiring
\[
\frac{\partial \chi^2}{\partial \rho_i} = \frac{\partial}{\partial \rho_i}\sum_j \frac{(\rho_i d_j-t_j)^2}{\varepsilon_j^2} = 0,
\]
where $j$ runs overs the data in the subset $i$, $d_j$, $\varepsilon_j$ and $t_j$ are respectively the $j$th data, its uncertainty and the associated theoretical prediction. We easily find
\[
\rho_i = \frac{\sum_j \frac{d_jt_j}{\varepsilon_j^2}}{\sum_j \frac{d_j^2}{\varepsilon_j^2}}.
\]
Finally, we shall require that $\rho_i$ does not lead to a normalisation bigger than the uncertainty $R_i$. This means that we shall constrain $\rho_i$ to verify
\[
1-R_i \le \rho_i \le 1+R_i.
\]

Before going to the result, one must point out that we have used here the latest CCFR data\footnote{They consist into a reanalysis of the 1997 data.} from 2001 \cite{Fleming:2000bg}. These data from U.K. Yang's thesis are used by adding the errors in quadrature and, in order to solve a discrepancy with the other data, we have also allowed an overall normalisation factor of at most 3\%.

\begin{table}
\begin{center}
\begin{tabular}{|c||c|c|}
\hline
Parameter & Value & Error \\
\hline
\hline
$A$ & 0.00876 & 0.00043 \\
$B$ & 0.0197 & 0.0035 \\
$C$ & 0.000 & 0.017 \\
\hline
$A_g$ & 0.258 &  0.032 \\
$B_g$ & -0.62 &  0.25 \\
\hline
$D_u$ & 0.378 & 0.030 \\
$D_d$ & 0.480 & 0.030 \\
$D_s$ & 0.000 & 0.013 \\
\hline
$\eta$ & 0.392 & 0.019 \\
\hline
$\gamma_u$ & 7.46 & 0.91 \\
$\gamma_d$ & 9.1 & 1.6 \\
\hline
$b_u$ & 3.625 & 0.016 \\
$b_d$ & 5.261 & 0.086 \\
$b$   & 6.67 & 0.27 \\
\hline
\hline
$N_u$ & 2.015 &   -   \\
$N_d$ & 1.723 &   -   \\
$C_g$ & 3.158 &   -   \\
\hline
\end{tabular}
\end{center}
\caption{Values of the fitted parameters in the parton distributions. The last three parameters are not fitted but are obtained from sum-rules.}\label{tab:xto1par}
\end{table}

\begin{landscape}
\begin{table}
\begin{center}
\begin{tabular}{|l|l|l|c|c||c|c||c|c||c|c|}
\hline
\multicolumn{5}{|c||}{Experiment information} & \multicolumn{2}{c||}{This fit} & \multicolumn{2}{c||}{CTEQ6 LO} & \multicolumn{2}{c|}{CTEQ6 NLO} \\
\hline
Quant.  & Colab. & Reference & Nb Pts & $\rho_i$ (\%) & $\chi^2$ & $\chi^2/nop$ & norm. & $\neg$ norm. & norm. & $\neg$ norm. \\
\hline
\hline
$F_2^p$ & BCDMS & PLB223(1989)485 &  167 &  -   &  154.607 & 0.926 & 5.303 & 5.303 & 2.652 & 2.652 \\
\cline{2-11}
        & E665  & PRD54(1996)3006 &   30 & 1.80 &   40.368 & 1.346 & 1.177 & 1.233 & 1.251 & 1.383 \\
\cline{2-11}
        & H1    & EPJC19(2001)269 &  126 &-1.50 &  129.673 & 1.029 & 1.516 & 1.626 & 1.077 & 1.122 \\
\cline{3-11}
        &       & EPJC21(2001)33  &   86 &  -   &   75.774 & 0.881 & 0.942 & 0.942 & 1.008 & 1.008 \\
\cline{3-11}
        &       & EPJC13(2000)609 &  130 &-1.50 &  117.682 & 0.905 & 1.612 & 1.962 & 0.882 & 1.032 \\
\cline{3-11}
        &       & NPB470(1996)3   &  156 &  -   &  104.206 & 0.668 & 0.835 & 0.835 & 0.658 & 0.658 \\
\cline{3-11}
        &       & NPB439(1995)471 &   90 &-4.50 &   49.499 & 0.550 & 0.597 & 0.901 & 0.574 & 0.737 \\
\cline{3-11}
        &       & NPB407(1993)515 &   21 &-8.00 &    6.233 & 0.297 & 0.289 & 0.466 & 0.287 & 0.401 \\
\cline{2-11}
        & NMC   & NPB483(1997)3   &   79 & 2.10 &  101.927 & 1.290 & 1.728 & 1.260 & 1.138 & 1.186 \\
\cline{2-11}
        & SLAC  & PLB282(1992)475 &   52 &  -   &   97.861 & 1.882 & 2.123 & 2.123 & 1.355 & 1.355 \\
\cline{2-11}
        & ZEUS  & EPJC21(2001)443 &  214 &  -   &  207.294 & 0.969 & 2.454 & 2.454 & 0.875 & 0.875 \\
\cline{3-11}
        &       & EPJC7(1999)609  &   12 &  -   &   11.297 & 0.941 & 0.744 & 0.744 & 1.259 & 1.259 \\
\cline{3-11}
        &       & ZPC72(1996)399  &  172 &  -   &  238.882 & 1.389 & 1.299 & 1.299 & 1.429 & 1.429 \\
\cline{3-11}
        &       & ZPC65(1995)379  &   56 & 2.00 &   27.477 & 0.491 & 0.495 & 0.415 & 0.453 & 0.470 \\
\cline{3-11}
        &       & ZPC69(1995)607  &    9 &-1.54 &   11.493 & 1.277 & 1.201 & 1.270 & 1.309 & 1.289 \\
\cline{3-11}
        &       & PLB316(1993)412 &   17 & 6.94 &    6.048 & 0.356 & 0.370 & 0.372 & 0.344 & 0.474 \\
\cline{2-11}
        & Total &                 & 1417 &      & 1380.321 & 0.974 & 1.864 & 1.900 & 1.150 & 1.187 \\
\hline
$F_2^d$ & BCDMS & PLB237(1989)592 &  154 &  -   &  127.941 & 0.831 & 1.546 & 1.546 & 0.903 & 0.903 \\
\cline{2-11}
        & E665  & PRD54(1996)3006 &   30 &  -   &   33.563 & 1.119 & 0.913 & 0.913 & 1.132 & 1.132 \\
\cline{2-11}
        & NMC   & NPB483(1997)3   &   79 & 1.00 &   90.330 & 1.143 & 1.438 & 1.131 & 0.969 & 1.071 \\
\cline{2-11}
        & SLAC  & SLAC-357(1990)  &   50 &  -   &   98.376 & 1.968 & 2.515 & 2.515 & 1.278 & 1.278 \\
\cline{2-11}
        & Total &                 &  313 &      &  350.210 & 1.119 & 1.613 & 1.536 & 1.002 & 1.027 \\
\hline
$F_2^{\nu N}$
        & CCFR  &UK. Yang's thesis&   65 & 3.00 &  165.512 & 2.546 & 3.118 & 4.570 & 3.523 & 6.135 \\
\hline
$xF_3^{\nu N}$
        & CCFR  & PRL79(1997)1213 &   76 &  -   &   42.066 & 0.554 & 0.658 & 0.658 & 1.252 & 1.252 \\
\hline
$F_2^n/F_2^p$
        &  NMC  & NPB371(1995)3   &   91 &  -   &  116.720 & 1.283 & 1.315 & 1.315 & 1.285 & 1.285 \\
\hline
\hline
Total   &       &                 & 1962 &      & 2054.830 & {\bf 1.047} & 1.794 & 1.855 & 1.215 & 1.333 \\
\hline
\end{tabular}
\end{center}
\caption{Fit results detailed experiment by experiment. For comparison we have added the predictions for CTEQ6 at leading and next-to-leading order (the NLO predictions are taken in the DIS scheme). In the comparison with CTEQ, the results are given with and without taking into account our normalisation factors.}\label{tab:xto1-chi2}
\end{table}
\end{landscape}

\section{Results of the DGLAP global fit}

\begin{figure}[ht]
\begin{center}
\includegraphics[scale=0.8]{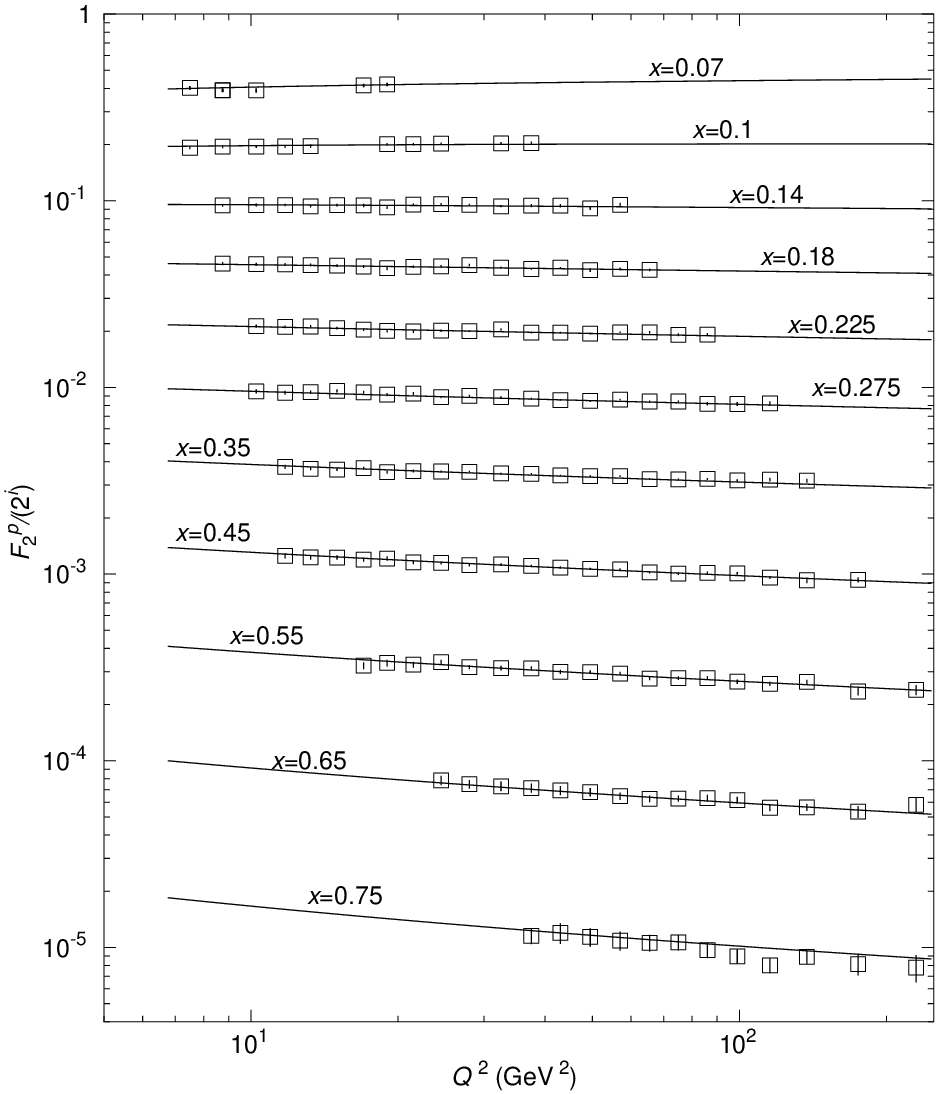}
\end{center}
\caption{DGLAP evolution results for BCDMS $F_2^p $ data ($i=0$ for the upper curve and is increased by 1 from one curve to the next one).}\label{fig:xto1-f2p-bcdms}
\end{figure}

\begin{figure}[ht]
\begin{center}
\includegraphics[scale=0.8]{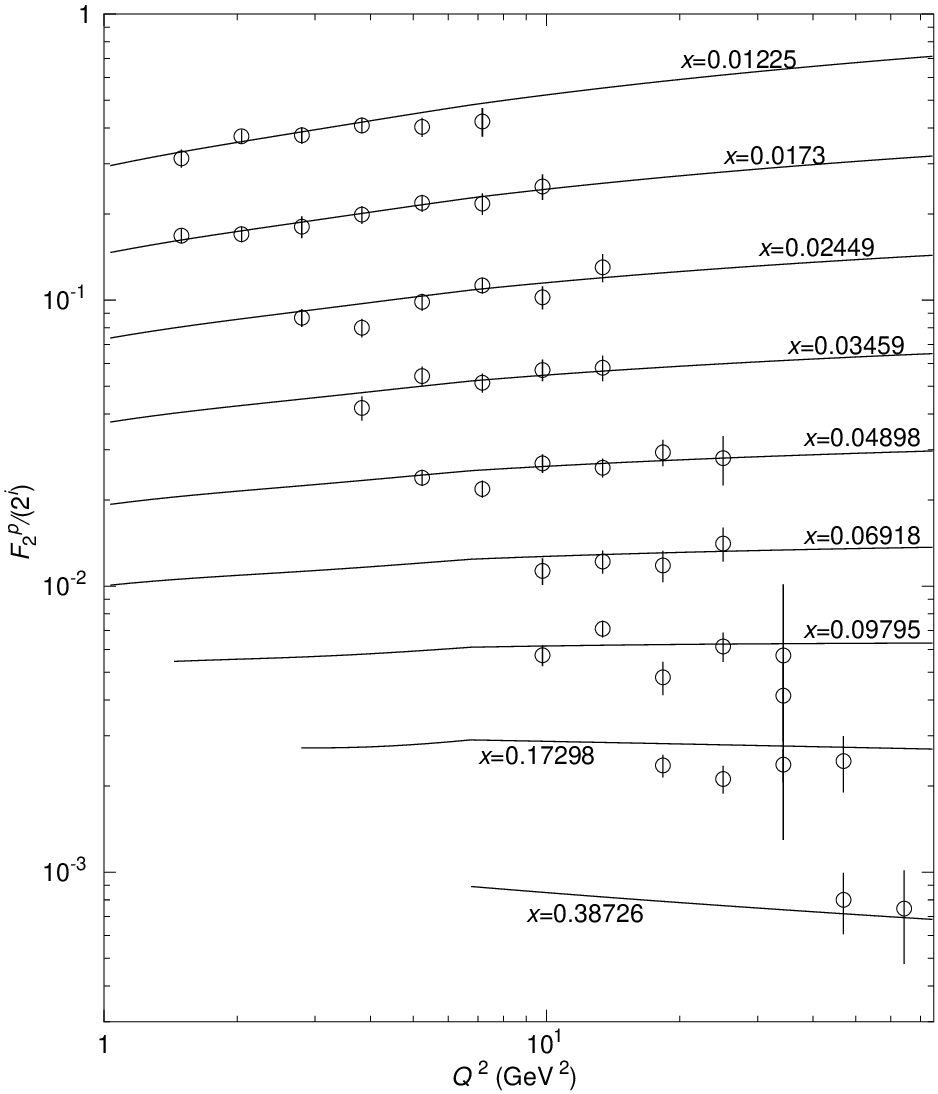}
\end{center}
\caption{DGLAP evolution results for E665 $F_2^p$ data. $i=0$ for the upper curve and is increased by 1 from one curve to the next one.}\label{fig:xto1-f2p-e665}
\end{figure}

\begin{figure}[ht]
\begin{center}
\includegraphics[scale=0.8]{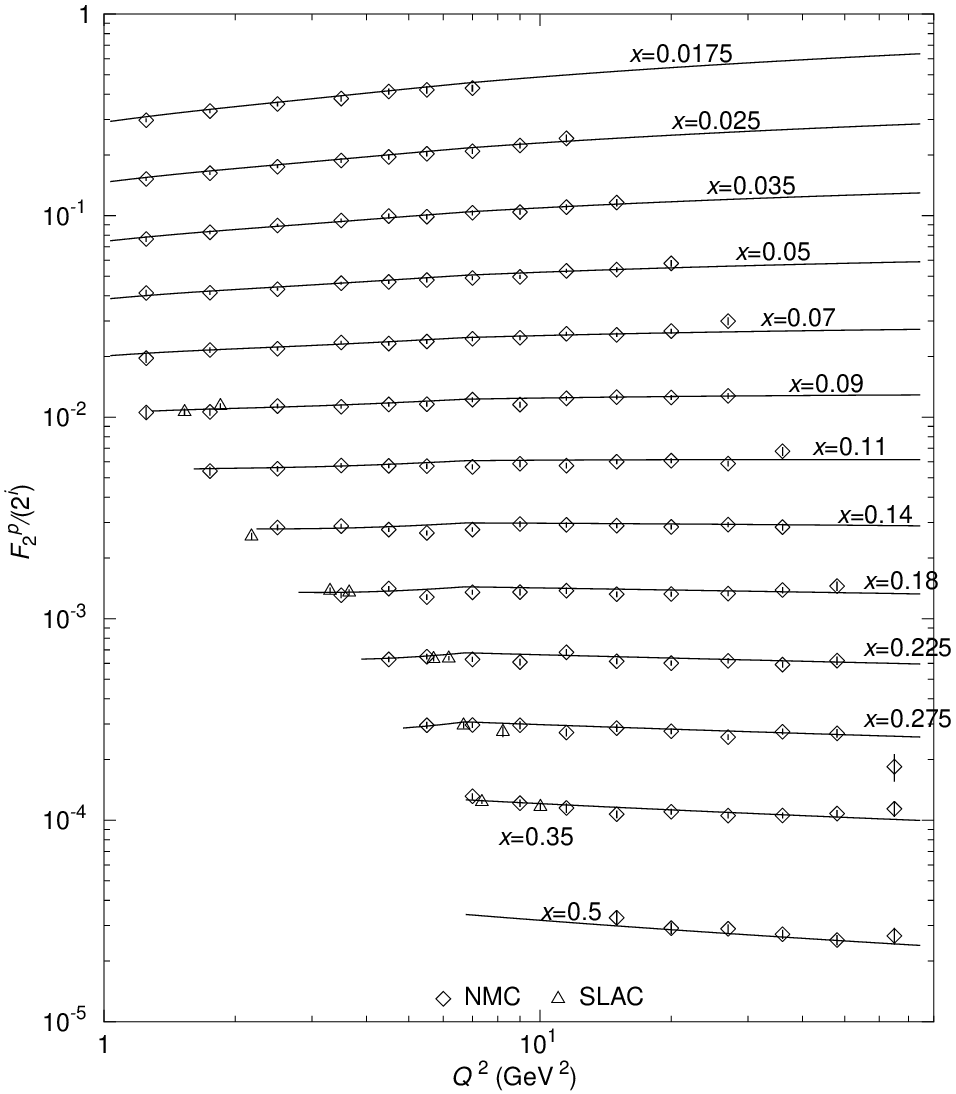}
\end{center}
\caption{DGLAP evolution results for NMC $F_2^p$ data (the SLAC data appearing in the NMC $Q^2$ bins have been added to the plot). $i=0$ for the upper curve and is increased by 1 from one curve to the next one.}\label{fig:xto1-f2p-nmc}
\end{figure}

\begin{figure}[ht]
\begin{center}
\includegraphics[scale=0.8]{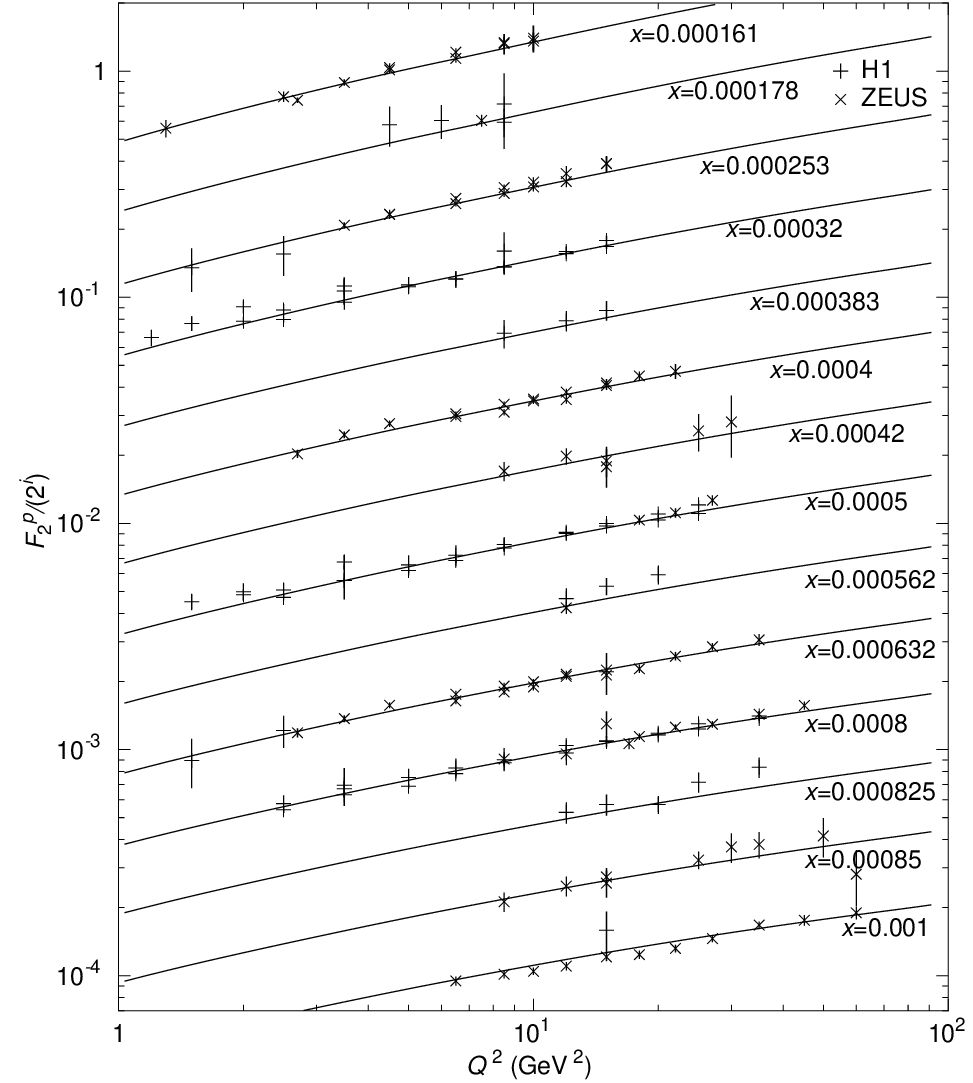}
\end{center}
\caption{DGLAP evolution results for HERA $F_2^p$ data ($x\le 0.001$). $i=0$ for the upper curve and is increased by 1 from one curve to the next one.}\label{fig:xto1-f2p-hera1}
\end{figure}

\begin{figure}[ht]
\begin{center}
\includegraphics[scale=0.8]{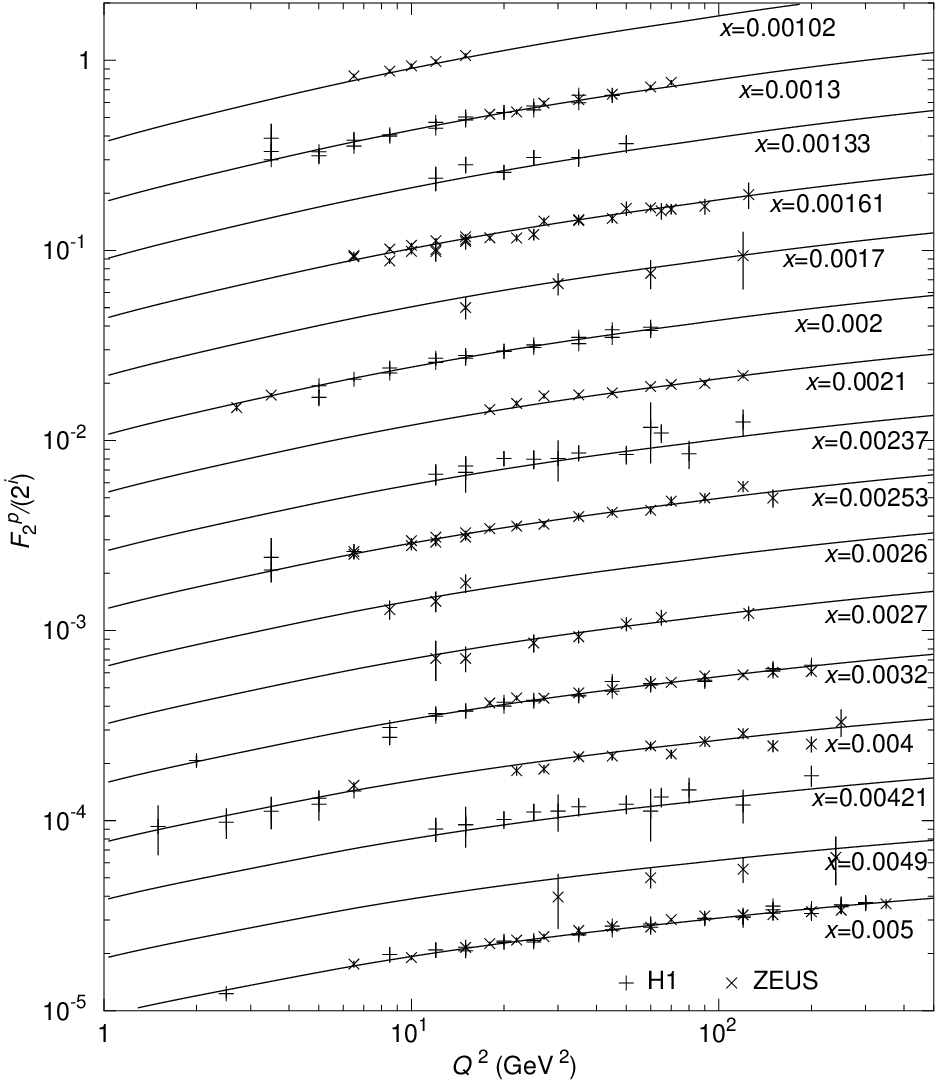}
\end{center}
\caption{DGLAP evolution results for HERA $F_2^p$ data ($0.001 < x\le 0.005$). $i=0$ for the upper curve and is increased by 1 from one curve to the next one.}\label{fig:xto1-f2p-hera2}
\end{figure}

\begin{figure}[ht]
\begin{center}
\includegraphics[scale=0.8]{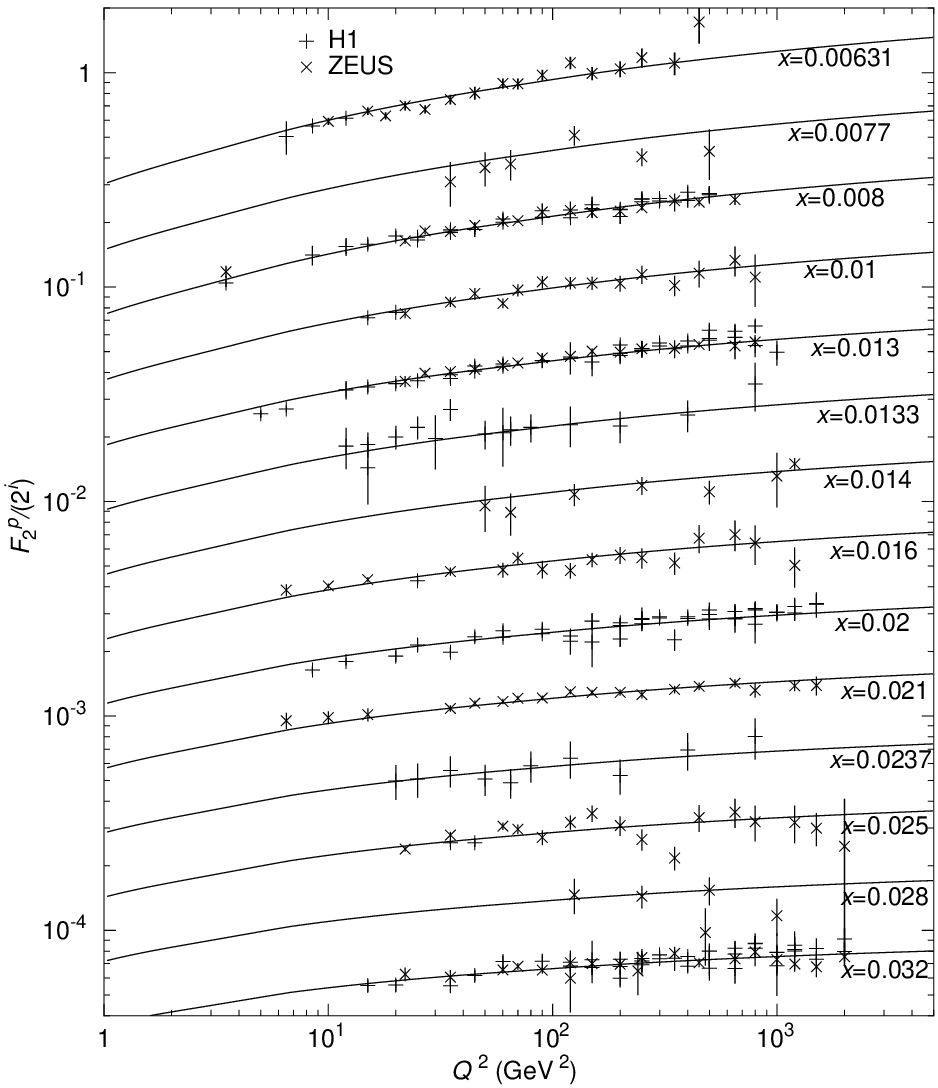}
\end{center}
\caption{DGLAP evolution results for HERA $F_2^p$ data ($0.005 < x \le 0.04$). $i=0$ for the upper curve and is increased by 1 from one curve to the next one.}\label{fig:xto1-f2p-hera3}
\end{figure}

\begin{figure}[ht]
\begin{center}
\includegraphics[scale=0.8]{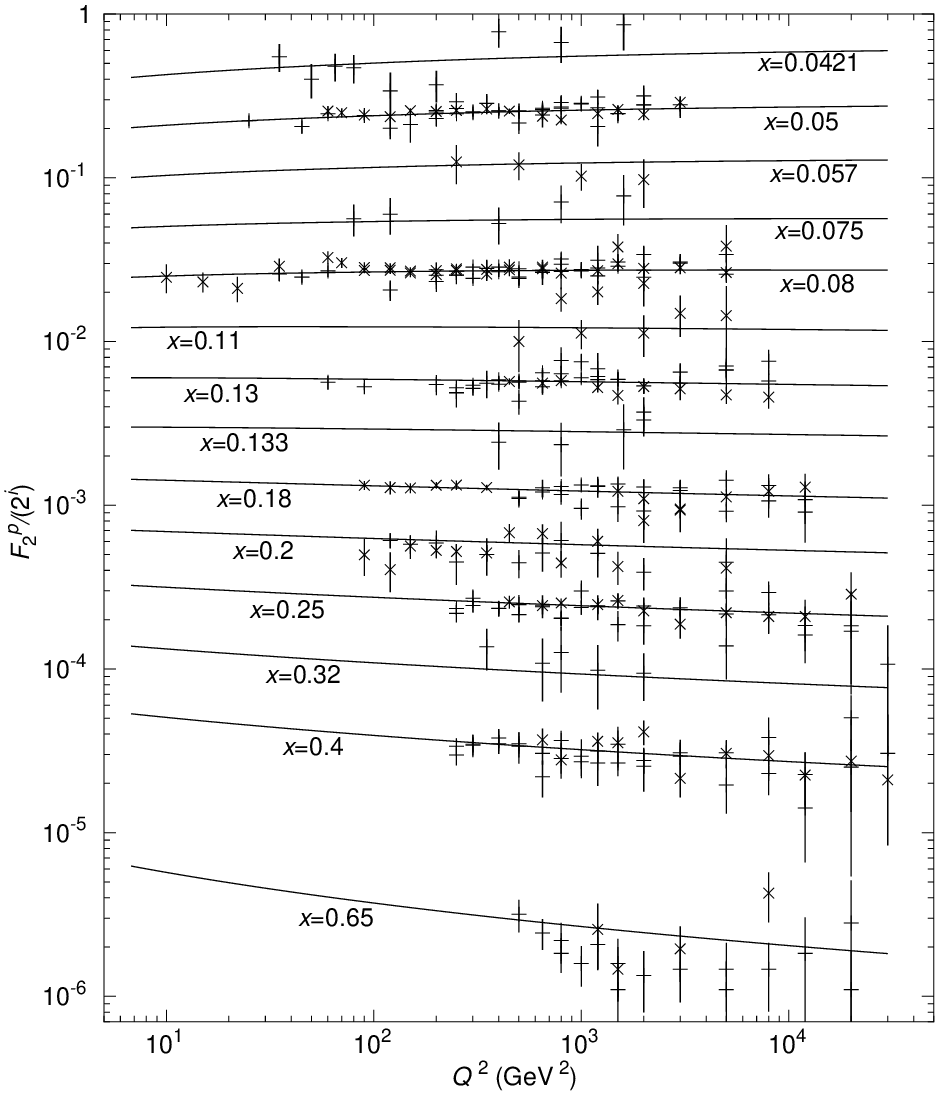}
\end{center}
\caption{DGLAP evolution results for HERA $F_2^p$ data ($0.04 < x$). $i=0$ for the upper curve and is increased by 1 from one curve to the next one.}\label{fig:xto1-f2p-hera-4}
\end{figure}

\begin{figure}[ht]
\begin{center}
\includegraphics[scale=0.8]{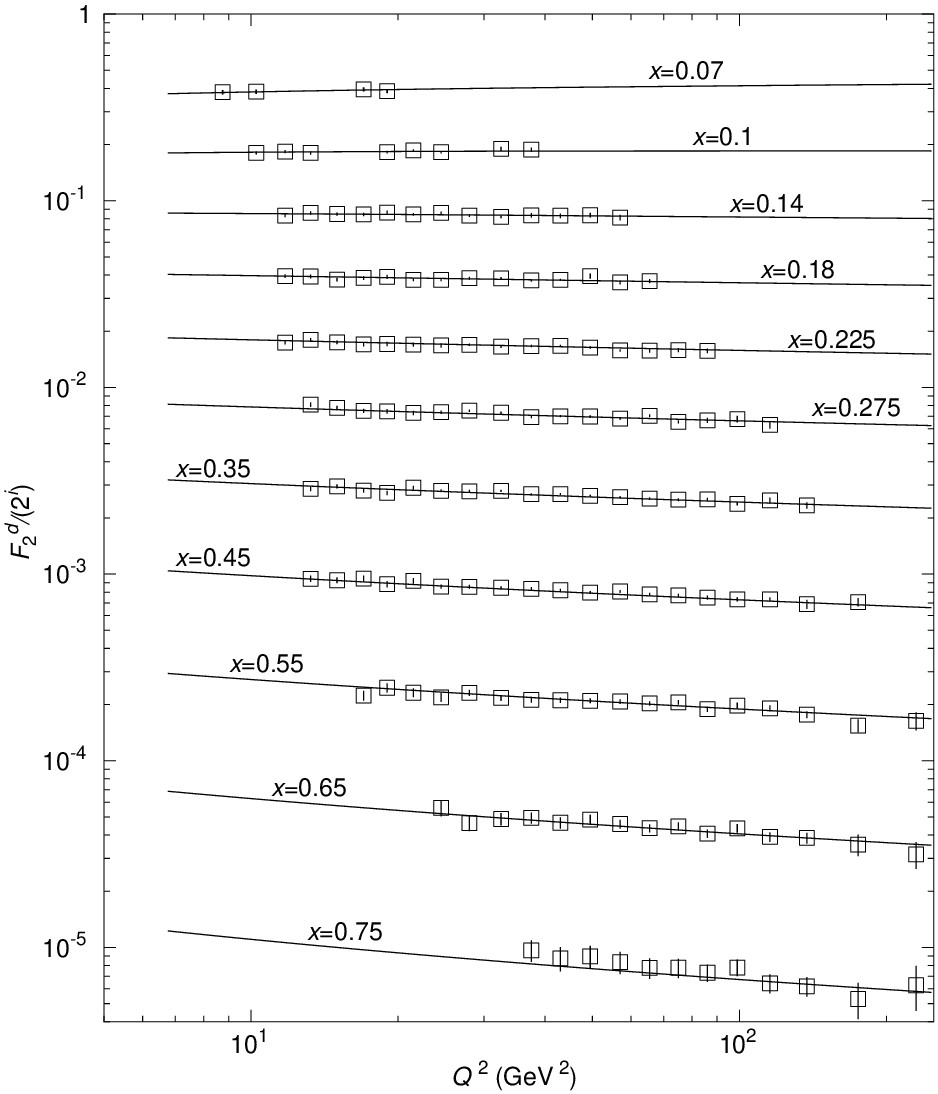}
\end{center}
\caption{DGLAP evolution results for BCDMS $F_2^d$ data. $i=0$ for the upper curve and is increased by 1 from one curve to the next one.}\label{fig:xto1-f2d-bcdms}
\end{figure}

\begin{figure}[ht]
\begin{center}
\includegraphics[scale=0.8]{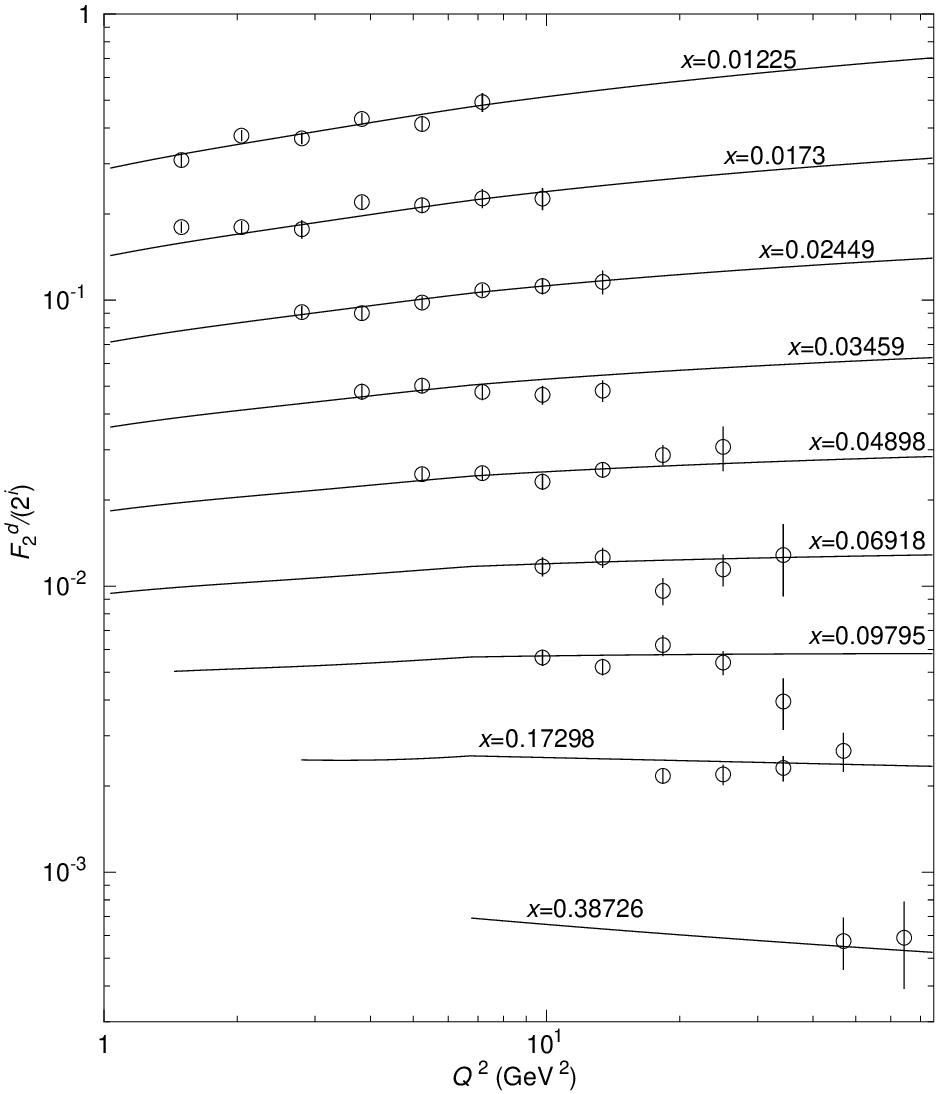}
\end{center}
\caption{DGLAP evolution results for E665 $F_2^d$ data. $i=0$ for the upper curve and is increased by 1 from one curve to the next one.}\label{fig:xto1-f2d-e665}
\end{figure}

\begin{figure}[ht]
\begin{center}
\includegraphics[scale=0.8]{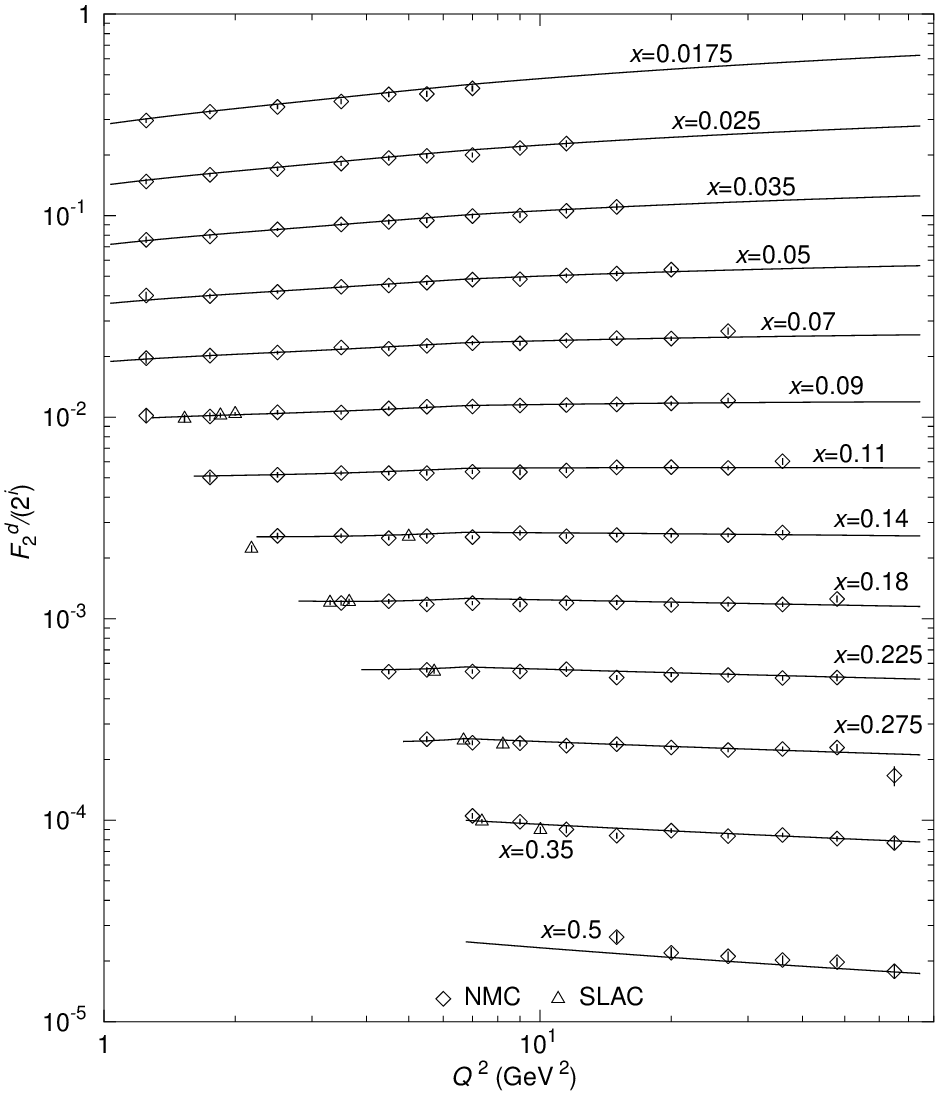}
\end{center}
\caption{DGLAP evolution results for NMC $F_2^d$ data (the SLAC data appearing in the NMC $Q^2$ bins have been added to the plot). $i=0$ for the upper curve and is increased by 1 from one curve to the next one.}\label{fig:xto1-f2d-nmd}
\end{figure}

\begin{figure}[ht]
\begin{center}
\includegraphics[scale=0.8]{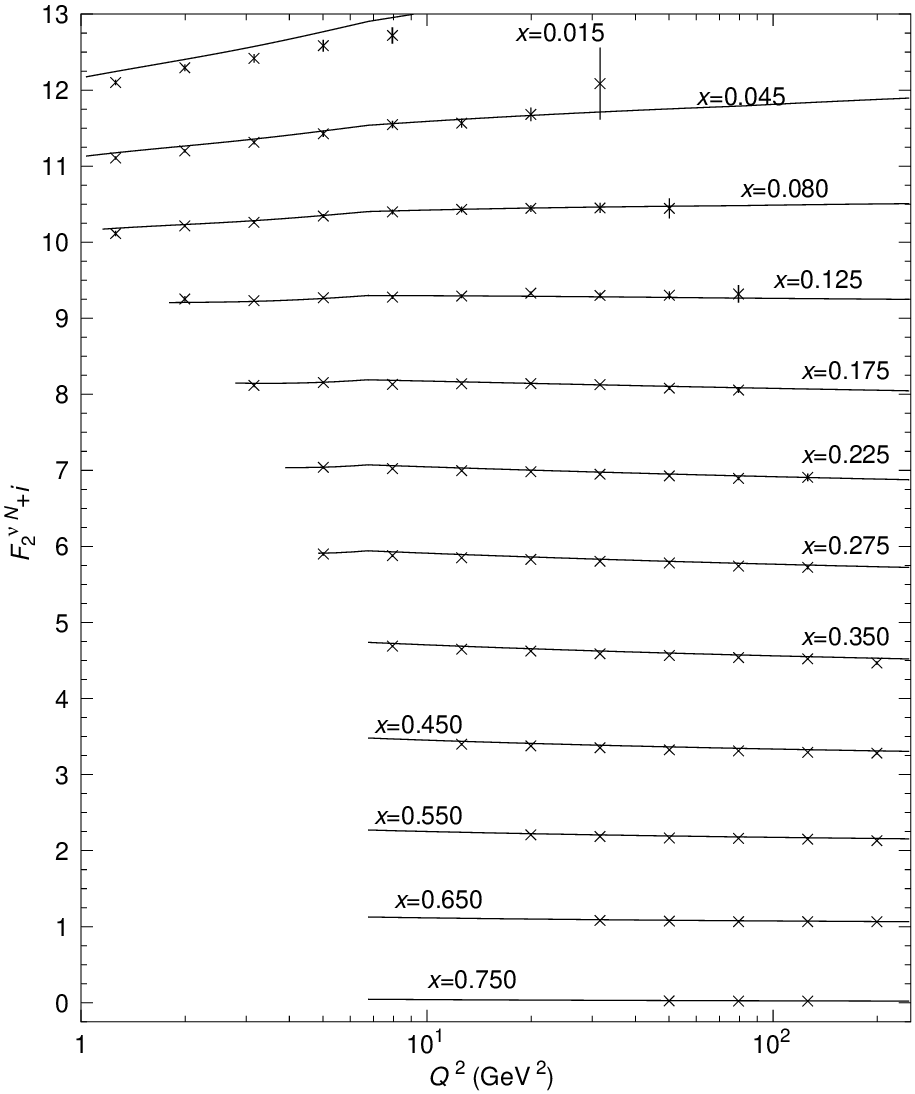}
\end{center}
\caption{DGLAP evolution results for CCFR $F_2^{\nu N}$ data. $i=0$ for the upper curve and is increased by 1 from one curve to the next one.}\label{fig:xto1-f2}
\end{figure}

\begin{figure}[ht]
\begin{center}
\includegraphics[scale=0.8]{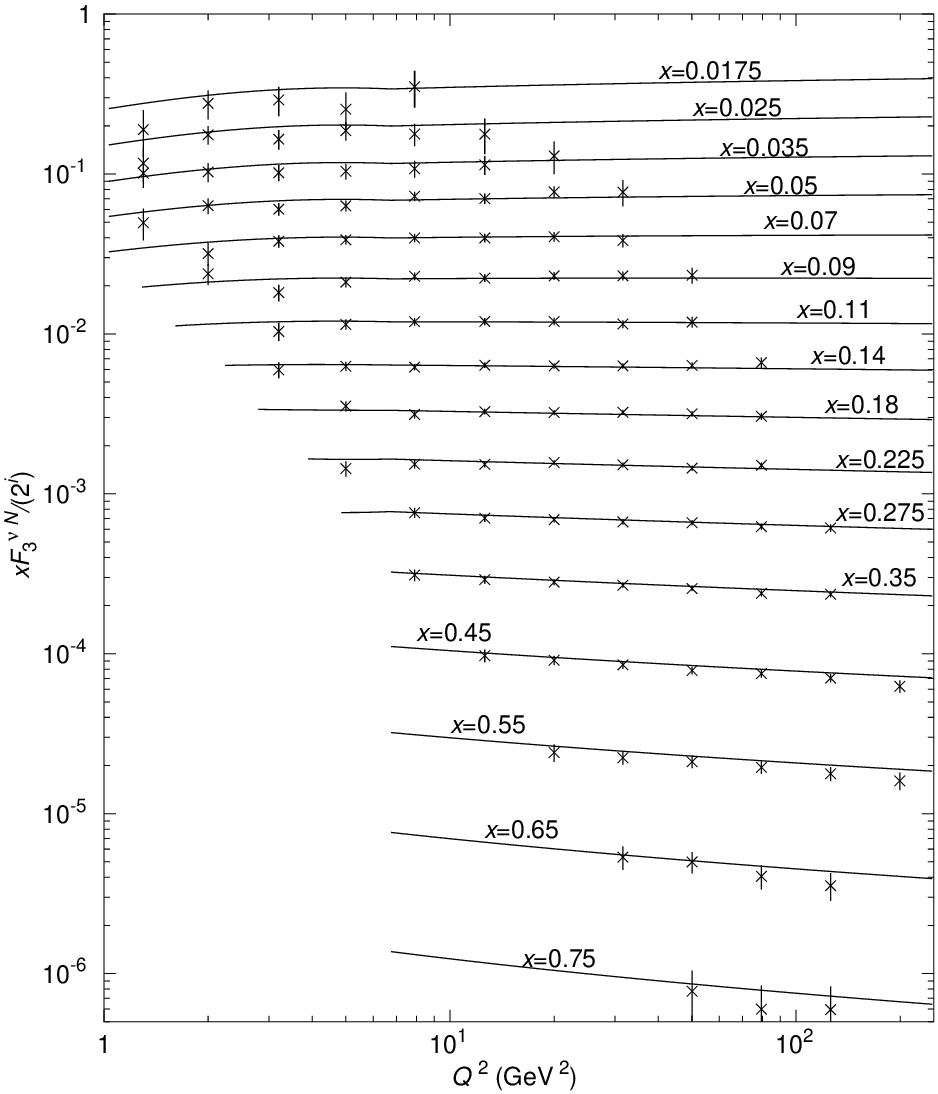}
\end{center}
\caption{DGLAP evolution results for CCFR $xF_3^{\nu N}$ data. $i=0$ for the upper curve and is increased by 1 from one curve to the next one.}\label{fig:xto1-f3nun}
\end{figure}

\begin{figure}[ht]
\begin{center}
\includegraphics[scale=0.8]{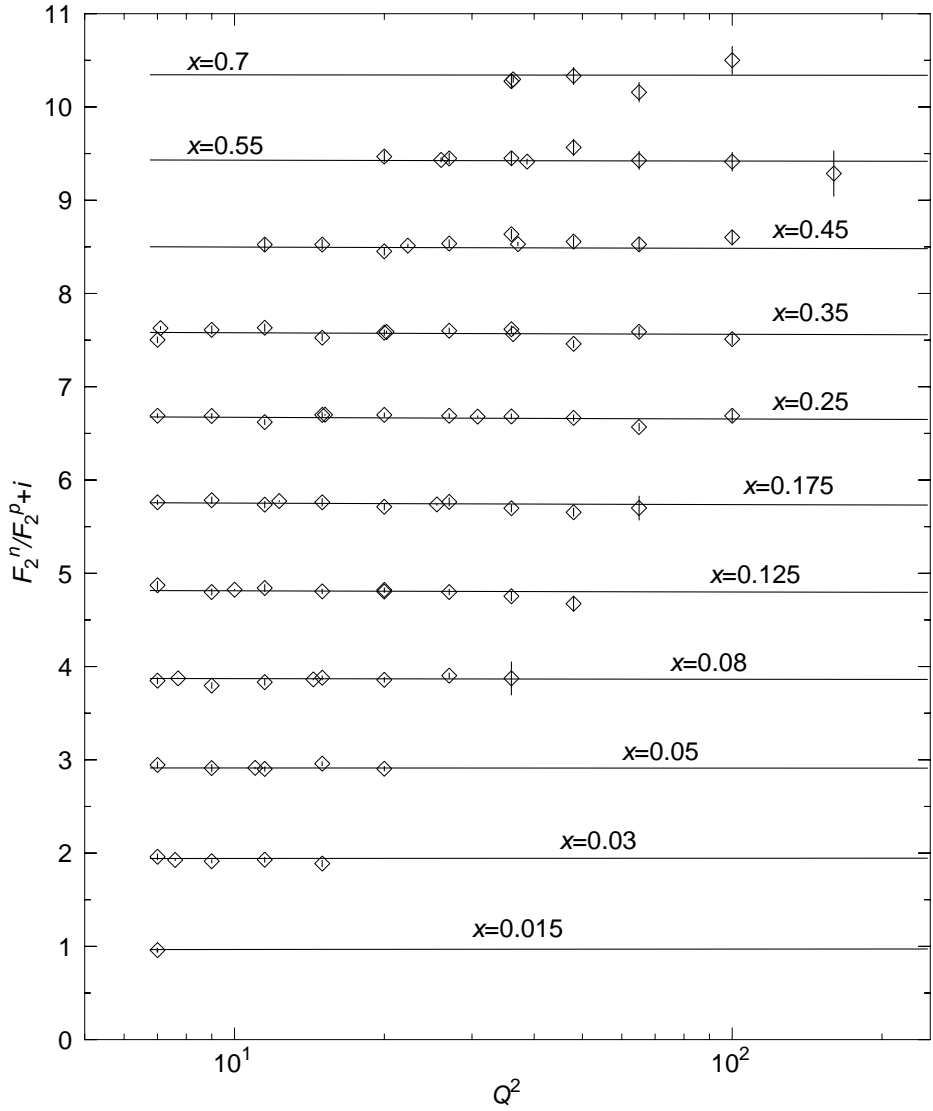}
\end{center}
\caption{DGLAP evolution results for NMC $F_2^n/F_2^p$ data. $i=0$ for the upper curve and is increased by 1 from one curve to the next one.}\label{fig:xto1-f2np}
\end{figure}

We have adjusted the 14 parameters $A$, $B$, $C$, $A_g$, $B_g$, $D_u$, $D_d$, $D_s$, $b$, $b_u$, $b_d$, $\gamma_u$, $\gamma_d$ and $\eta$ to the experimental data in the region
\begin{eqnarray*}
  Q^2 & \ge & 4m_c^2 = 6.76\,\text{GeV}^2,\\
W^2 & \ge & 12.5\,\text{GeV}^2.
\end{eqnarray*}
The second boundary is used to cut the region where higher-twists effects are expected to be large and we have adopted the same limit on $W^2$ as MRST. The values of the fitted parameters are presented in Table \ref{tab:xto1par} and the result, detailed experiment by experiment, is given in Table \ref{tab:xto1-chi2}. In addition, the curves resulting from our fit are presented for each experiment in Figures \ref{fig:xto1-f2p-bcdms} to \ref{fig:xto1-f2np}.

We can see from the parameter table that both the large-$x$ exponents and the reggeon intercept have acceptable values.

In order to evaluate the quality of our fit, we have also shown in Table \ref{tab:xto1-chi2} the CTEQ6 results at LO and at NLO (in the DIS scheme\footnote{The DIS scheme is the renormalisation scheme where, at any order, the quark coefficient function is $\delta(1-x)$ and the gluon coefficient function vanishes.}), with and without taking the normalisation factors into account\footnote{The CTEQ6 results are obtained by using the last CTEQ parton distributions to predict structure functions without any refit. Therefore, the LO and NLO results are just given for comparison.}. We see that the CCFR 2001 neutrino data probably need to be renormalised up and are still poorly reproduced. We can also see that, apart from the SLAC data, we obtain a very good description. This means that it would be a good idea to add a renormalisation factor of a few percents to the SLAC $F_2^p$ and $F_2^d$ data.

The correlation matrix for the parameters is presented in Table \ref{tab:xto1-corel}.

\begin{figure}
\begin{tabular}{cc}
\subfigure[]{\includegraphics[scale=0.8]{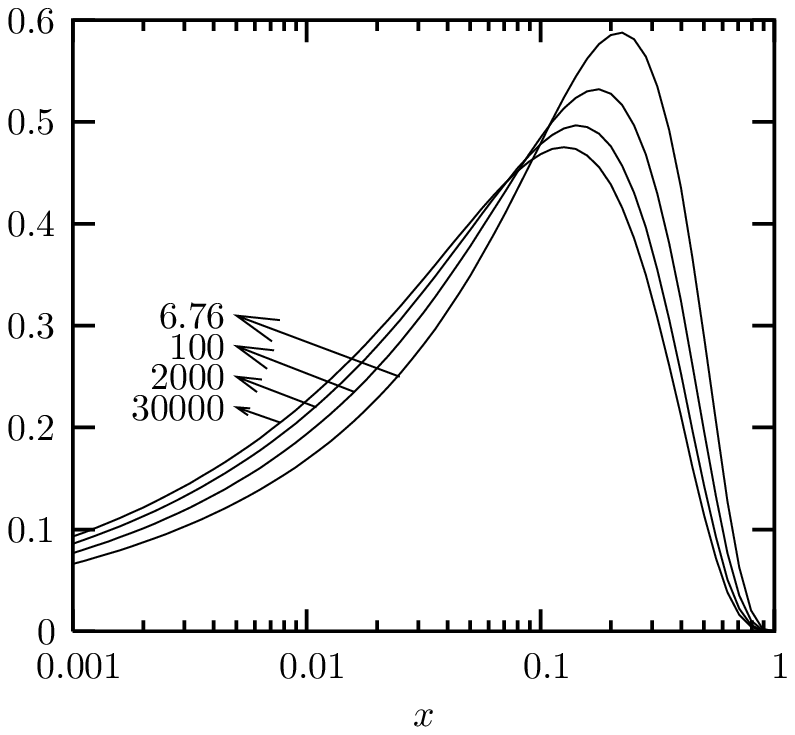}}&
\subfigure[]{\includegraphics[scale=0.8]{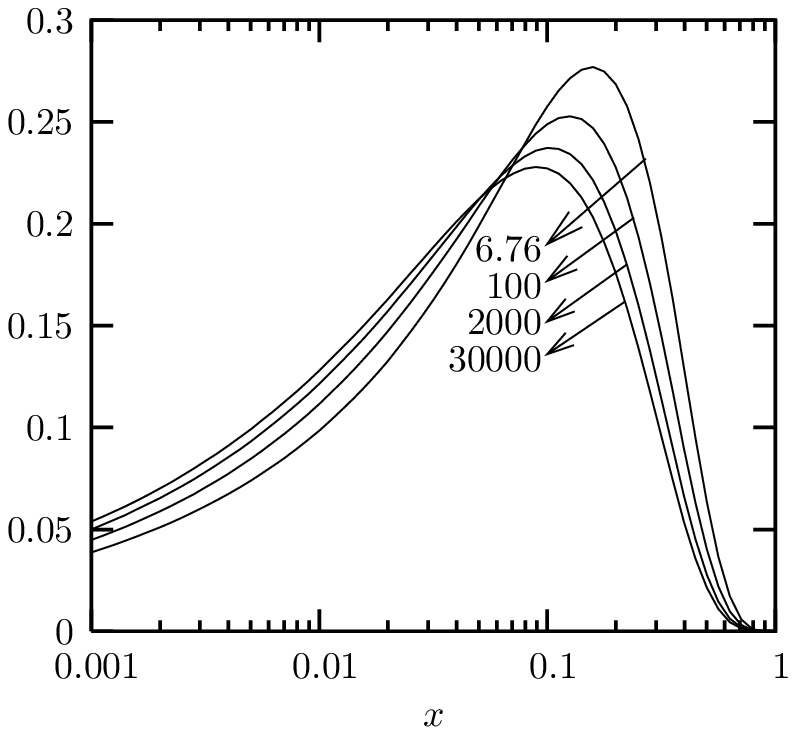}}\\
\subfigure[]{\includegraphics[scale=0.8]{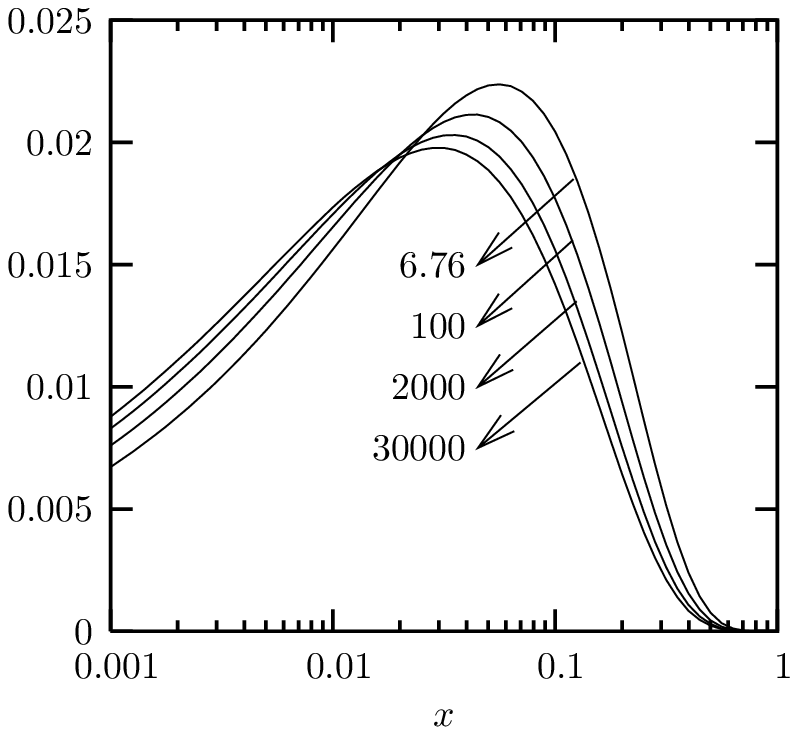}}&
\subfigure[]{\includegraphics[scale=0.8]{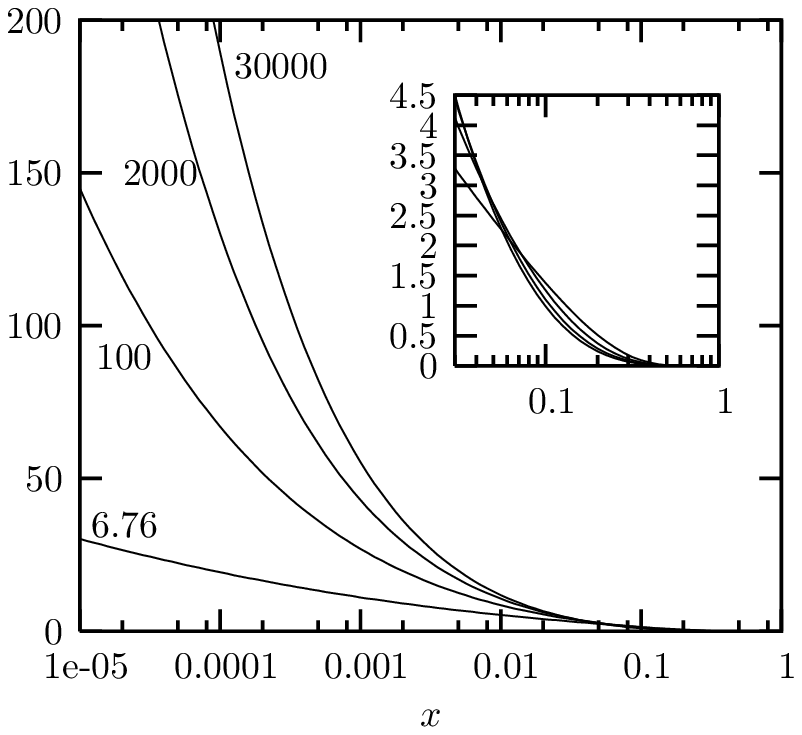}}
\end{tabular}
\caption{Typical momentum distributions inside the proton at various $Q^2$: (a) $u$ valence quarks, (b) $d$ valence quarks, (c) sea asymmetry $\bar{d}-\bar{u}$ and (d) gluon distribution.}\label{fig:xto1-q1}
\end{figure}

In Figure \ref{fig:xto1-q1}, we have shown some typical distributions and their $Q^2$ evolution. The $xu_V$ and $xd_V$ valence quarks distributions both present a peak around $x\approx 0.1-0.2$ and are, roughly speaking, within a factor 2. The sea asymmetry $\bar{d}-\bar{u}$ can be written in the following form:
\begin{eqnarray*}
x(\bar{d}-\bar{u}) & = & \frac{xu_V-xd_V-T_3}{2}\\
                   & = & (D_d-D_u) x^\eta (1-x)^b.
\end{eqnarray*}
This distribution has a maximum for
\[
x = \frac{\eta}{b+\eta} \approx \begin{cases} 
0.1 & \text{for }xu_V,\\
0.07 & \text{for }xd_V,\\
0.056 & \text{for }x(\bar{d}-\bar{u}).
\end{cases}
\]
The evolution in $Q^2$ of these three distribution shows the same behaviour: the peak is moved to smaller values of $x$ and tamed while its width grows.
We have also shown in Figure \ref{fig:xto1-q1} the gluon distribution which grows quickly with $Q^2$.

\begin{figure}
\begin{tabular}{cc}
\subfigure[]{\includegraphics[scale=0.8]{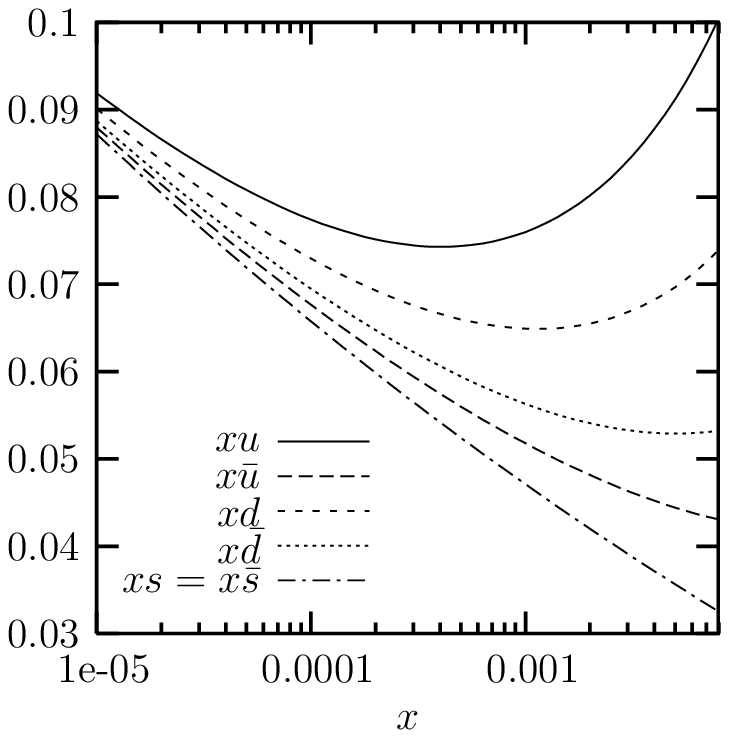}}&
\subfigure[]{\includegraphics[scale=0.8]{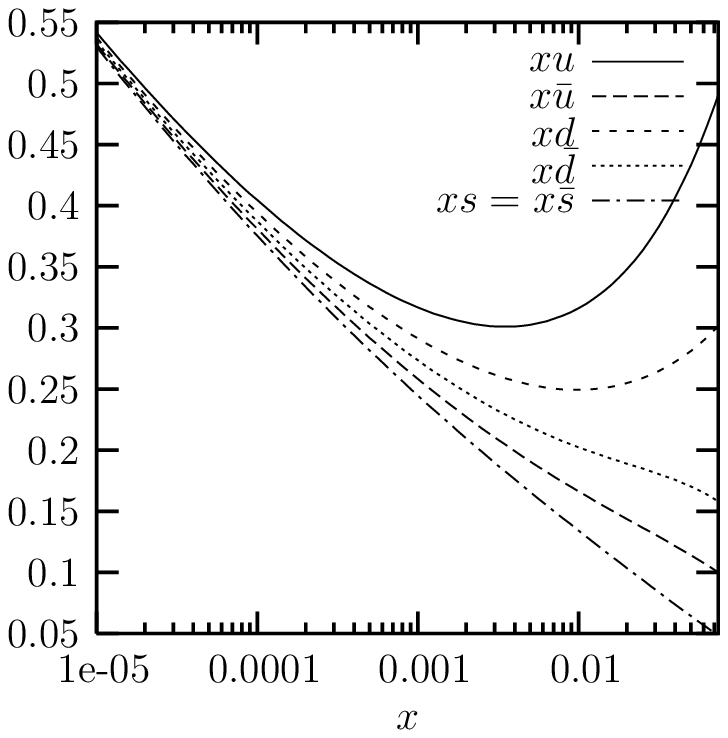}}\\
\subfigure[]{\includegraphics[scale=0.8]{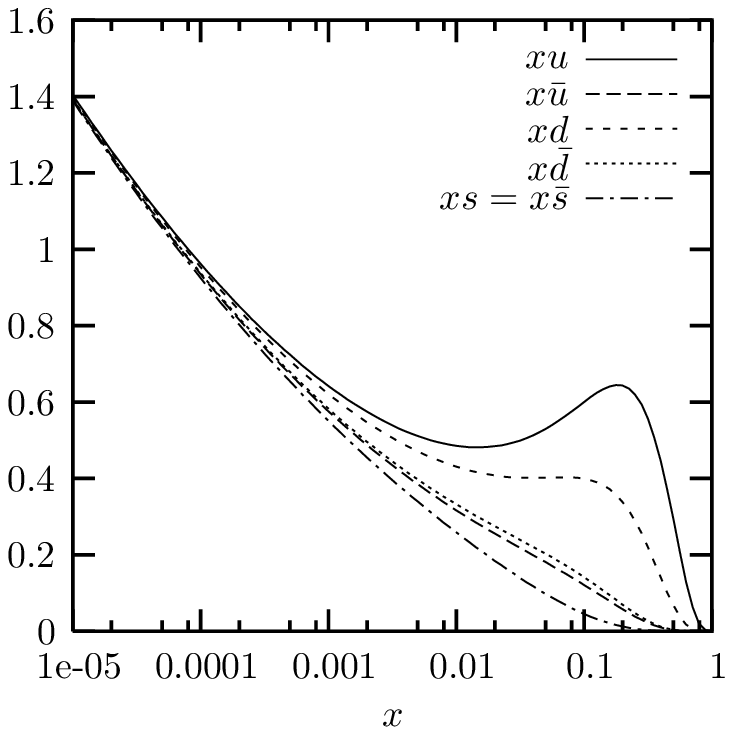}}&
\subfigure[]{\includegraphics[scale=0.8]{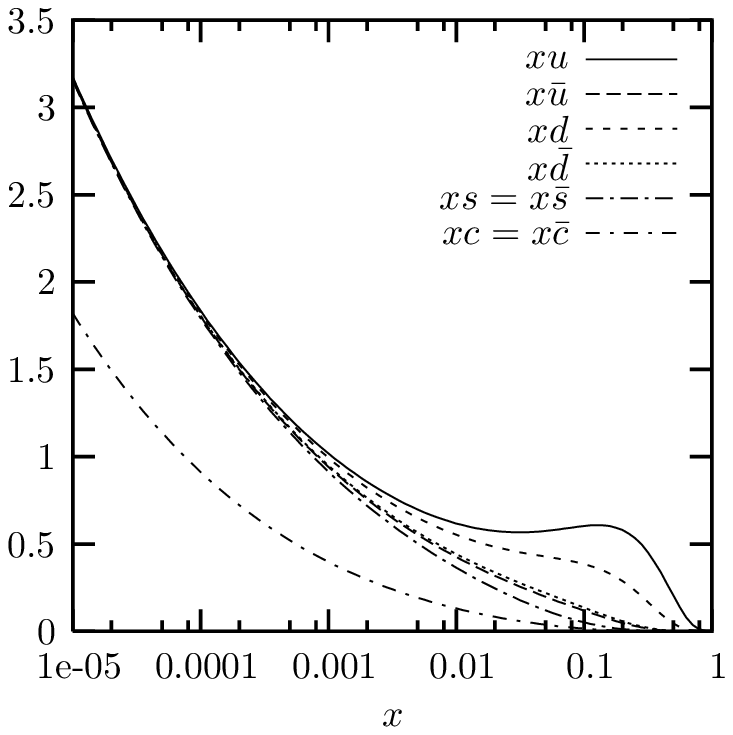}}\\
\subfigure[]{\includegraphics[scale=0.8]{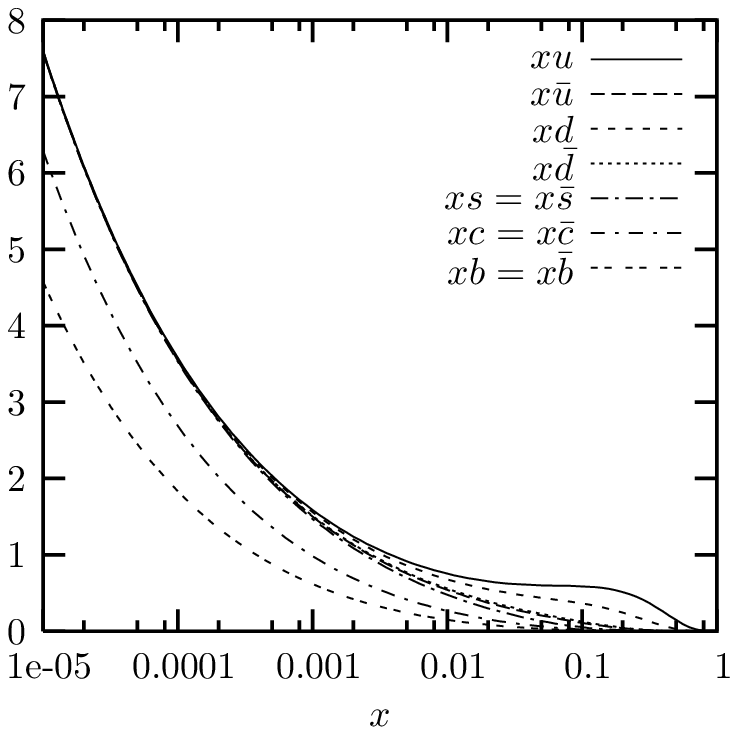}}&
\subfigure[]{\includegraphics[scale=0.8]{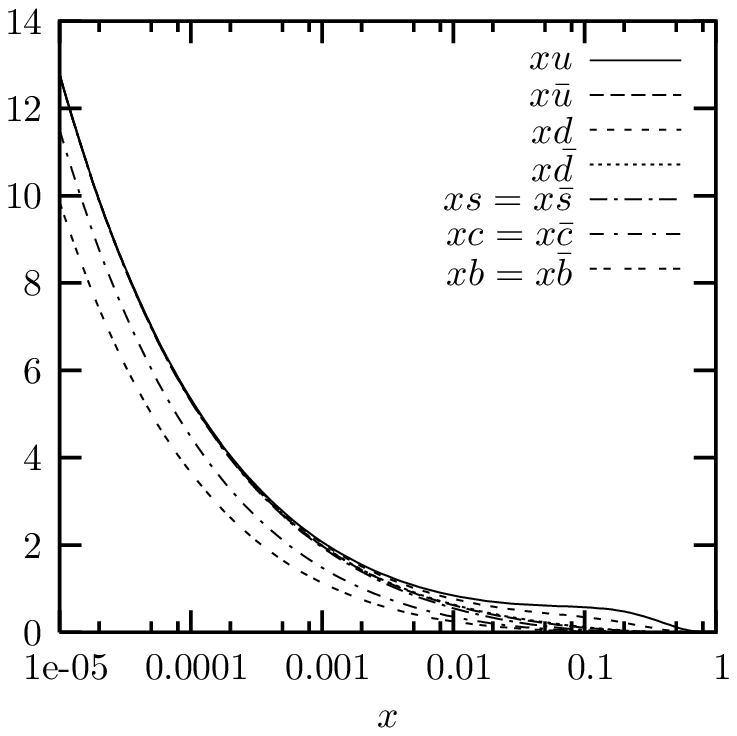}}
\end{tabular}
\caption{Quark distributions inside the proton at various $Q^2$: (a)~$Q^2=0.1$ GeV$^2$, (b)~$Q^2=1$ GeV$^2$, (c)~$Q^2=Q_0^2=4m_c^2$, (d)~$Q^2=4m_b^2$, (e)~$Q^2=2000$ GeV$^2$ and (f)~$Q^2=30000$ GeV$^2$. See \ref{sec:smallq2} for details concerning the parton distributions at small $Q^2$.}\label{fig:xto1-q2}
\end{figure}

The parton densities at various scales are plotted in Figure \ref{fig:xto1-q2}. First of all, when $Q^2=Q_0^2=4 m_c^2$, we have no charm or bottom quark and both quark and gluon distributions are described by Regge theory, more precisely by a triple-pole and a reggeon contribution. At higher virtualities, charm quarks are non-vanishing and, for $Q^2 > 4m_b^2$, we also have $b$ quarks. For $Q^2 > Q_0^2$, the parton distributions have an essential singularity at $j=1$.

Finally, we can estimate the uncertainty on the initial distributions in the following way: for the sea quarks or for the gluon, we have ($D=0$ for the gluon distribution)
\[
xq = \left[A \log^2(1/x) + B \log(1/x) + C + D x^\eta\right] (1-x)^{b}.
\]
If we assume that the uncertainties on the parameters are uncorrelated, we obtain easily
\begin{eqnarray*}
  (\delta xq)^2 & = & \left\{\log^4(1/x)\delta A^2 + \log^2(1/x) \delta B^2+\delta C^2 + \left[\delta D^2+\log^2(1/x)D^2 \delta \eta^2\right] x^{2\eta} \right. \\
                & + & \left.\left[A\log^2(1/x)+B\log(1/x)+C+D x^\eta\right]\log^2(1-x)\delta b^2\right\} (1-x)^{2b}.
\end{eqnarray*}
For the valence quarks, the initial distribution has the form
\[
xq_V = K x^\eta (1-x)^b (1+\gamma x)
\]
with $K$ fixed by quark number conservation, and we find that the uncertainty is
\[
(\delta xq)^2 = K^2 x^{2\eta} (1-x)^{2b} \left\{\left[\log^4(1/x)\delta \eta^2 +\log^2(1-x)\delta b^2\right] (1+\gamma x)^2 + x^2 \delta \gamma^2\right\}.
\]

\begin{figure}[ht]
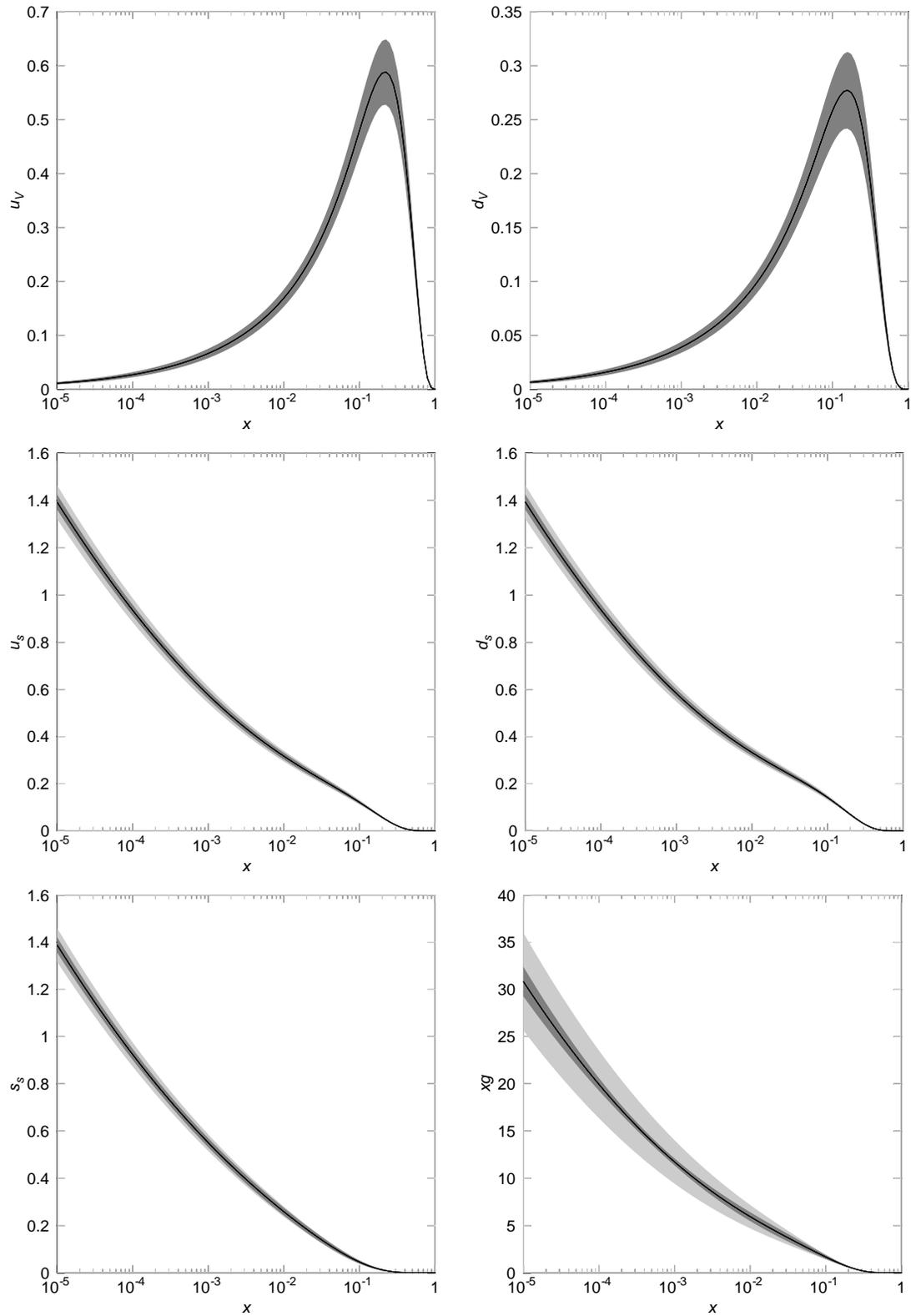

\begin{center}
\begin{tabular}{cc}
\includegraphics[scale=0.6]{figs/u-error.ps} &
\includegraphics[scale=0.6]{figs/d-error.ps} \\
\includegraphics[scale=0.6]{figs/us-error.ps} &
\includegraphics[scale=0.6]{figs/ds-error.ps} \\
\includegraphics[scale=0.6]{figs/ss-error.ps} &
\includegraphics[scale=0.6]{figs/g-error.ps}
\end{tabular}
\end{center}
\caption{Initial distributions with their uncertainties: the dark region represents the correlated uncertainties while the light one is obtained without taking into account the correlations between the parameters.}\label{fig:uncertainty}
\end{figure}

The resulting uncertainties on the initial distributions are shown in Figure \ref{fig:uncertainty}, where we have also plotted the uncertainties obtained by taking into account correlations between the parameters (see Table \ref{tab:xto1-corel}). We see that this ``traditional'' way of estimating errors leads to much smaller uncertainties than the joint consideration of forward and backward evolution obtained in \cite{Soyez:2003sr}.

\section{Regge theory at low $Q^2$}

\begin{table}
\begin{center}
\begin{tabular}{|l|l|l|c|c||c|c|}
\hline
\multicolumn{5}{|c||}{Experiment information} & \multicolumn{2}{c|}{This fit} \\
\hline
Quant.  & Colab. & Reference & Nb Pts & Norm. & $\chi^2$ & $\chi^2/nop$ \\
\hline
\hline
$F_2^p$       & E665 & PRD54(1996)3006  &  61 &  1.80 &  55.265 & 0.906 \\
\cline{2-7}
              & H1   & NPB439(1995)471  &   3 & -4.50 &   0.834 & 0.278 \\
\cline{3-7}
              &      & NPB470(1996)3    &  37 &   -   &  15.518 & 0.419 \\
\cline{3-7}
              &      & NPB497(1996)3    &  44 & -3.00 &  35.660 & 0.810 \\
\cline{3-7}
              &      & EPJC21(2001)33   &  47 &   -   &  61.721 & 1.313 \\
\cline{2-7}
              & NMC  & NPB483(1997)3    &  67 &  2.10 &  46.473 & 0.694 \\
\cline{2-7}
              & SLAC & PLB282(1992)475  &  94 &   -   & 102.417 & 1.090 \\
\cline{2-7}
              & ZEUS & ZPC69(1995)607   &  14 &  1.54 &  29.452 & 2.104 \\
\cline{3-7}
              &      & ZPC72(1996)399   &  16 &   -   &  11.339 & 0.709 \\
\cline{3-7}
              &      & PLB407(1997)432  &  34 &   -   &  10.753 & 0.316 \\
\cline{3-7}
              &      & EPJC7(1999)609   &  32 &   -   &  34.541 & 1.079 \\
\cline{3-7}
              &      & EPJC12(2000)35   &  70 &   -   &  86.120 & 1.230 \\
\cline{3-7}
              &      & EPJ21(2001)443   &  28 &   -   &  48.034 & 1.716 \\
\cline{2-7}
              & Total&                  & 547 &   -   & 538.127 & 0.984 \\
\hline
$F_2^d$       & E665 & PRD54(1996)3006  &  61 &   -   &  76.977 & 1.262 \\
\cline{2-7}
              & NMC  & NPB483(1997)3    &  67 &  1.00 &  40.064 & 0.598 \\
\cline{2-7}
              & SLAC & SLAC-357(1990)   &  98 &   -   &  86.511 & 0.883 \\
\cline{3-7}
              &      & PRD49(1994)5641  &   1 &  0.89 &   0.000 & 0.000 \\
\cline{2-7}
              & Total&                  & 227 &   -   & 203.552 & 0.897 \\
\hline
$F_2^{\nu N}$ & CCFR & UK. Yang's thesis&  19 &  3.00 &  63.159 & 3.325 \\
\hline
$xF_3^{\nu N}$& CCFR & PRL79(1997)1213  &  35 &   -   &  39.201 & 1.120 \\
\hline
$F_2^n/F_2^p$ & NMC  & NPB371(1995)3    & 120 &   -   & 101.445 & 0.845 \\
\hline
\hline
Total         &      &                  & 948 &   -   & 945.485 & 0.997 \\
\hline
\end{tabular}
\end{center}
\caption{Result of the small-$Q^2$ fit detailed experiment by experiment.}\label{tab:lowchi2}
\end{table}

\begin{table}
\begin{center}
\begin{tabular}{|l|c|c|}
\hline
Parameter             & Value & Error \\
\hline
$a_{\cB}$             & 15.05 & 1.80  \\
$Q_{\cA}^2$           &  6.37 & 2.72  \\
$Q_{\cB}^2$           & 1.885 & 0.522 \\
$Q_{\cC}^2$           & 10.00 & 8.70  \\
$Q_{\cD_u}^2$         & 0.437 & 1.07  \\
$Q_{\cD_d}^2$         &  5.19 & 3.11  \\
$Q_{\cb}^2$           &  3.64 & 1.87  \\
$Q_{\cb_u}^2$         &  5.10 & 2.38  \\
$Q_{\cb_d}^2$         &  87.9 & 15.8  \\
$\varepsilon_{\cA}$   & 1.002 & 0.328 \\
$\varepsilon_{\cB}$   & 0.581 & 0.110 \\
$\varepsilon_{\cC}$   & 18.19 & 4.37  \\
$\varepsilon_{\cD_u}$ & 0.340 & 0.157 \\
$\varepsilon_{\cD_d}$ & 1.618 & 0.615 \\
$\varepsilon_{\cb}$   & 1.916 & 0.453 \\
$\varepsilon_{\cb_u}$ & 2.558 & 0.593 \\
$\varepsilon_{\cb_d}$ & 10.00 & 6.94  \\
\hline
\end{tabular}
\end{center}
\caption{Value of the parameters with their errors for the low-$Q^2$ fit. The scales are given in GeV$^2$.}\label{tab:lowparam}
\end{table}

\subsection{Motivation}

If DGLAP evolution gives the behaviour of the parton distributions at large $Q^2$, we expect soft physics to be described by Regge theory. In other words, Regge theory should not only be able to describe the initial DGLAP condition but also the structure functions for $0 \le Q^2 \le Q_0^2$. In this section, we shall therefore try to extend the parton distribution functions at low values of $Q^2$. Note that in this region, we cannot use the DGLAP equation anymore. In addition, if we want to use Regge theory, we must still restrict ourselves to the high-energy domain. Hence we keep the constraint
\[
W^2 \ge 12.5\text{ GeV}^2
\]
but allow $Q^2$ to be in the region
\[
0 \le Q^2 \le 6.76\text{ GeV}^2.
\]

\subsection{Small-$Q^2$ parametrisation}\label{sec:smallq2}

If we want to use Regge theory in the small-$Q^2$ region, we need to parametrise the parton distribution functions. We shall use the same expressions as in \eqref{eq:initdistrib} with an additional $Q^2$ dependence. However, if we want to consider the extension down to $Q^2=0$, we know that we should use the Regge variable $\nu=\frac{Q^2}{2x}$ instead of $x$. This means that we shall use the following distributions:
\begin{eqnarray}
xu_V(\nu,Q^2) & = & \frac{2}{N_u^*} (2\nu)^{-\eta}\left[1+\cg_u(Q^2) \frac{Q^2}{2\nu}\right]\left(1-\frac{Q^2}{2\nu}\right)^{\cb_u(Q^2)},\\
xd_V(\nu,Q^2) & = & \frac{1}{N_d^*} (2\nu)^{-\eta}\left[1+\cg_d(Q^2) \frac{Q^2}{2\nu}\right]\left(1-\frac{Q^2}{2\nu}\right)^{\cb_d(Q^2)},\nonumber\\
xu_s(\nu,Q^2) & = & \left\{\cA(Q^2)\left[\log(2\nu)-\cB(Q^2)\right]^2+\cC(Q^2)+\cD_u(Q^2) (2\nu)^{-\eta} \right\}\left(1-\frac{Q^2}{2\nu}\right)^{\cb(Q^2)},\label{eq:lowq2distrib}\nonumber\\
xd_s(\nu,Q^2) & = & \left\{\cA(Q^2)\left[\log(2\nu)-\cB(Q^2)\right]^2+\cC(Q^2)+\cD_d(Q^2) (2\nu)^{-\eta} \right\}\left(1-\frac{Q^2}{2\nu}\right)^{\cb(Q^2)},\nonumber\\
xs_s(\nu,Q^2) & = & \left\{\cA(Q^2)\left[\log(2\nu)-\cB(Q^2)\right]^2+\cC(Q^2)+\cD_s(Q^2) (2\nu)^{-\eta} \right\}\left(1-\frac{Q^2}{2\nu}\right)^{\cb(Q^2)},\nonumber
\end{eqnarray}
where, once again, $N_u$ and $N_d$ are constrained by quark number conservation. We shall require that the parameters in these distributions match the initial distribution taken for DGLAP evolution at 6.76 GeV$^2$. Using parametrisations of the form\footnote{Since $\cD_s$ already vanishes at $Q^2=Q_0^2$, we have set it to zero in the whole small-$Q^2$ region.}
\begin{eqnarray*}
\phi(Q^2) = a_\phi Q^2 \left(\frac{Q_\phi^2}{Q^2+Q_\phi^2}\right)^{\varepsilon_\phi}& \qquad\text{ for } & \phi=\cA,\cC,\cD_u,\cD_d, \cb,\cb_u,\cb_d\\
\cB(Q^2) = a_{\cB} \left(\frac{Q^2}{Q^2+Q_{\cB}^2}\right)^{\varepsilon_{\cB}}+a_{\cB}^*& &\\
\cg_i(Q^2) = \cg_i(Q_0^2)& \qquad\text{ for } & i=u,d,\\
\cD_s = 0&&
\end{eqnarray*}
and constraining the parameters $a_\cA$, $a_\cB^*$, $a_\cC$, $a_{\cD_u}$, $a_{\cD_d}$, $a_{\cD_s}$, $a_{\cb}$, $a_{\cb_u}$ and $a_{\cb_d}$ with the DGLAP initial condition at $Q^2=Q_0^2=6.76$ GeV$^2$, we are left with 17 parameters: $a_\cB$, $Q_\cA^2$, $Q_\cB^2$, $Q_\cC^2$, $Q_{\cD_u}^2$, $Q_{\cD_d}^2$, $Q_\cb^2$, $Q_{\cb_u}^2$, $Q_{\cb_d}^2$, $\varepsilon_\cA$, $\varepsilon_\cB$, $\varepsilon_\cC$, $\varepsilon_{\cD_u}$, $\varepsilon_{\cD_d}$, $\varepsilon_\cb$, $\varepsilon_{\cb_u}$, $\varepsilon_{\cb_d}$. The expressions obtained once the constrained have been imposed are the following:
\begin{eqnarray*}
\cA(Q^2)   & = & A\frac{Q^2}{Q_0^2}*\left(\frac{Q_0^2+Q_{\cA}^2}{Q^2+Q_{\cA}^2}\right)^{\varepsilon_{\cA}}\\
\cB(Q^2)   & = & a_{\cB}\left[\left(\frac{Q^2}{Q^2+Q_{\cB}^2}\right)^{\varepsilon_{\cB}}-\left(\frac{Q_0^2}{Q_0^2+Q_{\cB}^2}\right)^{\varepsilon_{\cB}}\right]+\log(Q_0^2)-\frac{B}{2A} \\
\cC(Q^2)   & = & \left(C-\frac{B^2}{4A}\right)\frac{Q^2}{Q_0^2}\left(\frac{Q_0^2+Q_{\cC}^2}{Q^2+Q_{\cC}^2}\right)^{\varepsilon_{\cC}} \\
\cD_u(Q^2) & = & D_u\frac{Q^2}{Q_0^2}\left(Q_0^2\right)^\eta\left(\frac{Q_0^2+Q_{\cD_u}^2}{Q^2+Q_{\cD_u}^2}\right)^{\varepsilon_{\cD_u}} \\
\cD_d(Q^2) & = & D_d\frac{Q^2}{Q_0^2}\left(Q_0^2\right)^\eta\left(\frac{Q_0^2+Q_{\cD_d}^2}{Q^2+Q_{\cD_d}^2}\right)^{\varepsilon_{\cD_d}} \\
\cD_s(Q^2) & = & 0 \\
\cb(Q^2)   & = & b  \frac{Q^2}{Q_0^2}*\left(\frac{Q_0^2+Q_{\cb  }^2}{Q^2+Q_{\cb  }^2}\right)^{\varepsilon_{\cb}}\\
\cb_u(Q^2) & = & b_u\frac{Q^2}{Q_0^2}*\left(\frac{Q_0^2+Q_{\cb_u}^2}{Q^2+Q_{\cb_u}^2}\right)^{\varepsilon_{\cb_u}}\\
\cb_d(Q^2) & = & b_d\frac{Q^2}{Q_0^2}*\left(\frac{Q_0^2+Q_{\cb_d}^2}{Q^2+Q_{\cb_d}^2}\right)^{\varepsilon_{\cb_d}}\\
\gamma_u(Q^2) & = & \gamma_u \\
\gamma_d(Q^2) & = & \gamma_d
\end{eqnarray*}

\subsection{Dataset and systematic errors}

In the small-$Q^2$ region ($W^2\ge 12.5$ GeV$^2$, $Q^2\le 6.76$ GeV$^2$), we shall fit the same quantities as previously and the data coming from the same collaborations:
\begin{itemize}
\item $F_2^p$: H1 \cite{Ahmed:1995fd,Aid:1996au,Adloff:1997mf,Adloff:2000qk}, ZEUS \cite{Derrick:1995ef,Derrick:1996hn,Breitweg:1997hz,Breitweg:1998dz,Breitweg:1999ad,Chekanov:2001qu}, NMC \cite{Arneodo:1996qe}, E665 \cite{Adams:1996gu}, SLAC \cite{Whitlow:1991uw},
\item $F_2^d$: BCDMS, NMC \cite{Arneodo:1996qe}, E665 \cite{Adams:1996gu}, SLAC \cite{Dasu:1993vk},
\item $F_2^n/F_2^p$: NMC \cite{Amaudruz:1991nw},
\item $F_2^{\nu N}$ and $xF_3^{\nu N}$: CCFR \cite{Fleming:2000bg}.
\end{itemize}

Concerning the treatment of the systematic errors, we have used the same correction factors as the ones obtained in the DGLAP global fit for the papers containing data in both the small- and the large-$Q^2$ region and leave this factor free for the papers containing only data at small $Q^2$.

\subsection{Results}

The parametrisations described above have been fitted to the 948 data in the small-$Q^2$ region using MINUIT. The results of this fit, experiments by experiments, together with the parameter values is presented in Tables \ref{tab:lowchi2} and \ref{tab:lowparam}. We see that, apart from the ZEUS 1995 data and the CCFR $F_2$ data, we obtain a good description of the structure functions in the low-$Q^2$ region. The $\chi^2$ per data point is quite good, considering that we have applied Regge theory to quite a large region as compared to usual approaches \cite{Donnachie:2001xx,Jenkovszky:1991fs,Desgrolard:kv,Desgrolard:2001bu,Cudell:2002ej,Soyez:2004pz}. The poor description of the ZEUS 1995 data may come from the fact that the systematic uncertainties have been fixed in the DGLAP fit.

The results for the structure functions in the low-$Q^2$ region are shown in Figures \ref{fig:xto1-f2p-bcdms} to \ref{fig:xto1-f2np} together with the large-$Q^2$ results. In the small-$Q^2$ region, the curves are only drawn in the fitted region ($W^2>12.5$ GeV$^2$). We see that the experimental measurements are well reproduced. 

Finally, the curves in Figure \ref{fig:xto1-q2} show the parton distribution functions obtained at small $Q^2$. It is interesting to notice that valence quarks are large at small $Q^2$ and $W^2\approx 12.5$ GeV$^2$ which, in this case, correspond to small values of $x$. For example at $Q^2=0.1$ GeV$^2$ and $x=0.008$, we have $x\bar{u} \approx 0.043$, $x\bar{d} \approx 0.053$, $xu_V \approx 0.057$ and $xd_V \approx 0.019$.

\section{DGLAP vs. Regge at high $Q^2$}

\subsection{Motivation}

As we have shown in \cite{Soyez:2003sr}, we can consider that Regge theory also applies at large $Q^2$. In these conditions, since $Q^2$-dependent singularities are forbidden in Regge theory, we expect a triple-pole behaviour at all values of $Q^2$. The unphysical essential singularity generated by DGLAP evolution should therefore be considered as a numerical approximation to a triple-pole pomeron at small $x$. 

Given these considerations, we have shown \cite{Soyez:2003sr} that, using both forward and backward evolution, it is possible to describe the small-$x$ experimental data with parton distribution functions of the form
\[
A(Q^2)\log^2(1/x) + B(Q^2)\log(1/x) + C(Q^2) + D(Q^2) x^\eta
\]
where the $Q^2$-dependent couplings are extracted from the DGLAP evolution equation.

\subsection{Parametrisation and uncertainties}

In this QCD global fit, we would like to test if it is still possible to consider the result of the evolution as an approximation to a triple-pole. To achieve this task, we shall fit the parton distribution functions at each value of $Q^2$ with the following form
\begin{eqnarray*}
xu_V & = & \frac{2}{N_u^*} x^\eta(1+\gamma_u x)(1-x)^{b_u},\\
xd_V & = & \frac{1}{N_d^*} x^\eta(1+\gamma_d x)(1-x)^{b_d},\\
xu_s & = & \left[A\log^2(1/x)+B\log(1/x)+C+D_u x^\eta \right](1-x)^b,\\
xd_s & = & \left[A\log^2(1/x)+B\log(1/x)+C+D_d x^\eta \right](1-x)^b,\\
xs_s & = & N_s\left[A\log^2(1/x)+B\log(1/x)+C+D_s x^\eta \right](1-x)^b,\\
xc_s & = & N_c\left[A\log^2(1/x)+B\log(1/x)+C+D_c x^\eta \right](1-x)^b,\\
xb_s & = & N_b\left[A\log^2(1/x)+B\log(1/x)+C+D_b x^\eta \right](1-x)^b.\\
xg   & = & \left[A_g\log^2(1/x)+B_g\log(1/x)+C_g^*\right](1-x)^{b+1},
\end{eqnarray*}

Before performing this fit, we shall determine an uncertainty on the parton distribution. Since we want to show that the DGLAP evolution generates an essential singularity which mimics a triple-pole behaviour, we should estimate the error introduced by the evolution. Since we have used LO DGLAP evolution, we estimate that the errors are of the order of the NLO corrections:
\[
q_{\text{NLO}}(x,Q^2) \approx (1\pm \alpha_s(Q^2)) q_{LO}(x,Q^2).
\]
If we require that the intimal parton distribution at $Q^2=Q_0^2$ remains fixed, this leads to
\[
q_{\text{NLO}}^{\text{norm}}(x,Q^2) \approx \frac{1\pm \alpha_s(Q^2)}{1\pm \alpha_s(Q_0^2)} q_{LO}(x,Q^2),
\]
or, keeping only the leading term in the strong coupling constant,
\[
\Delta q(x,Q^2) = \left|\alpha_s(Q_0^2)-\alpha_s(Q^2)\right| q_{LO}(x,Q^2).
\]
In addition, we shall take into account the fact that, at small $Q^2$, there may also be higher-twist corrections. If we assume\footnote{This value reproduces errors comparable with those obtained from our uncertainties estimation. In addition, the higher-twist term is relevant for middle-range values of $Q^2$ while the subleading corrections are important at large $Q^2$.} these are at 5\% at $Q^2=Q_0^2$, we shall finally consider
\[
\Delta q(x,Q^2) = \left[\left|\alpha_s(Q_0^2)-\alpha_s(Q^2)\right|+\frac{0.05\:Q_0^2}{(1-x)Q^2}\right] q_{LO}(x,Q^2).
\]
Finally, additional powers of $1-x$ are expected to describe the large-$x$ behaviour of the parton distributions at large $Q^2$, hence we shall only consider the region $10^{-5}\le x \le 0.1$ for the sea quarks and the gluons. For the case of the valence quarks, we hope that the factor $(1+\gamma x)$ is sufficient to reproduce the distribution for $10^{-5}\le x \le 1$.

\subsection{Results}

\begin{figure}[ht]\label{fig:ess-triple}
\begin{center}
\includegraphics{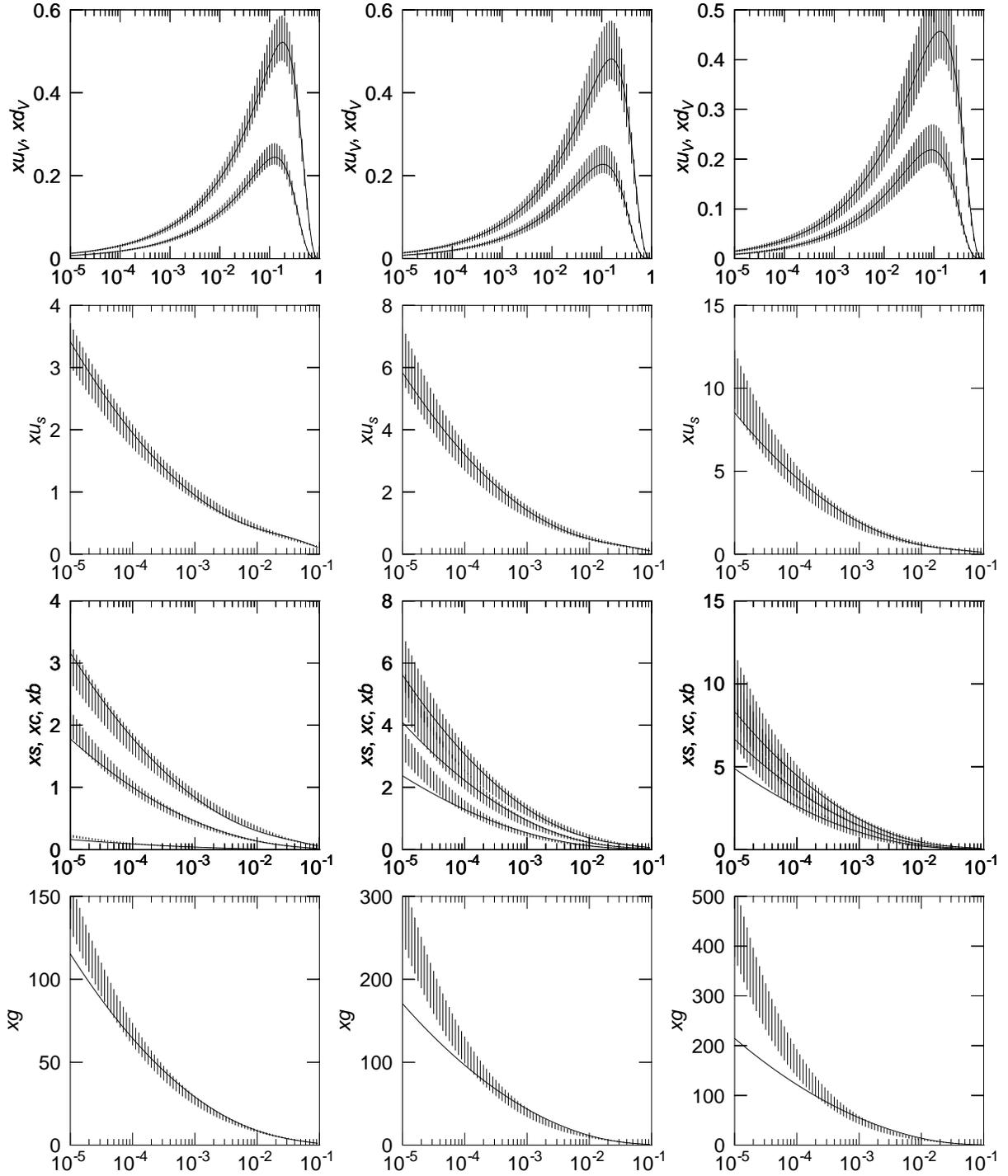}
\end{center}
\caption{Triple-pole pomeron fit to the parton distribution functions obtained from DGLAP evolution. The first column shows distributions at $Q^2=100$ GeV$^2$, the second corresponds to $Q^2=1000$ GeV$^2$ and the third to $Q^2=10000$ GeV$^2$.}
\end{figure}

\begin{figure}[ht]\label{fig:formfact}
\begin{center}
\includegraphics{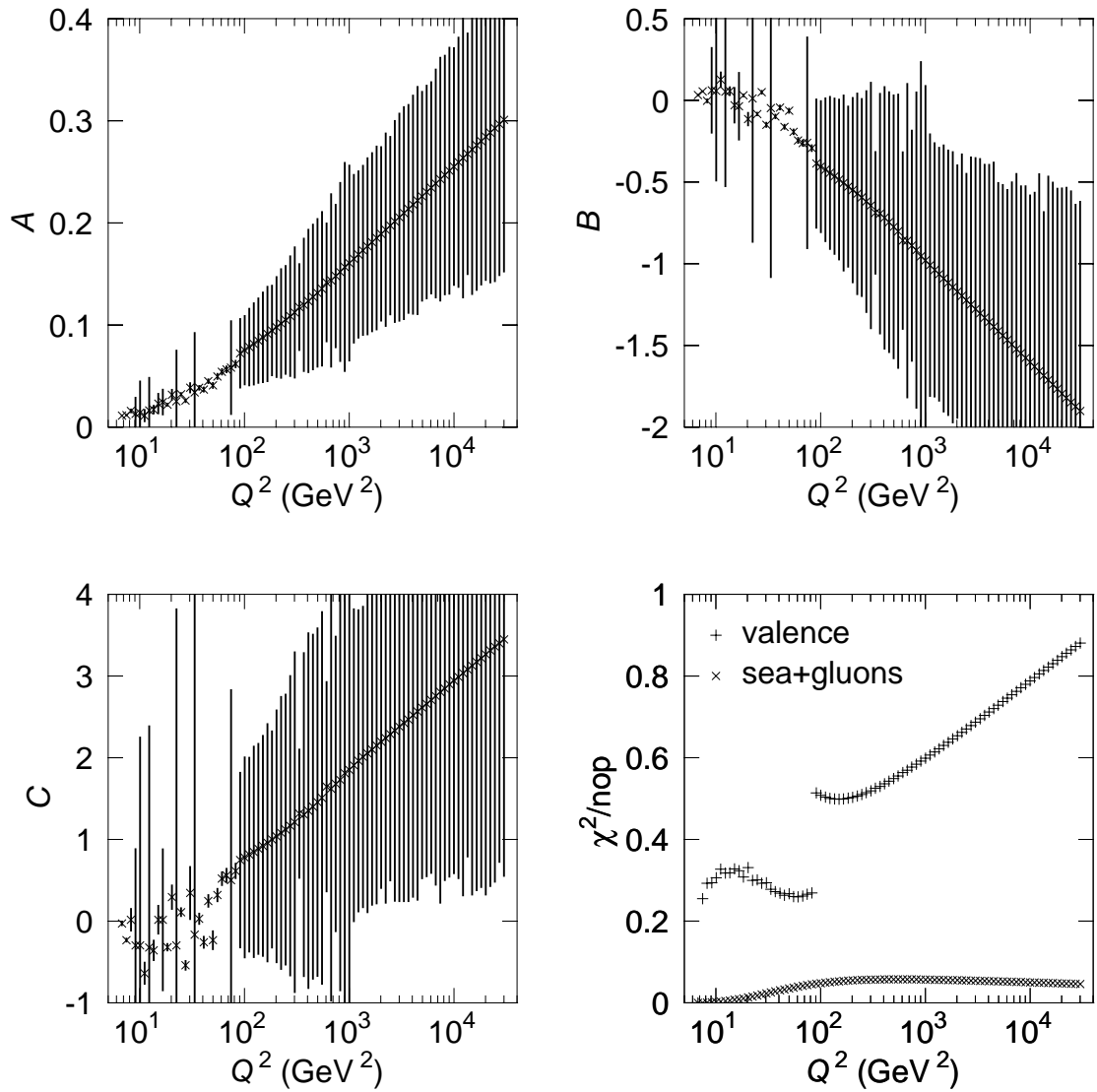}
\end{center}
\caption{Triple-pole form factors for $F_2^p$ at large $Q^2$ presented together with the result of the fit of a triple pole to the large-$Q^2$ parton densities/.}
\end{figure}

To perform the fit, we have taken the parton distributions with their estimated uncertainties in 80 points regularly spaced in $\log(x)$. Since the parametrisations for the valence quarks and the sea quarks have disjoint parameters, we have performed two different fits at each $Q^2$. The result of these are presented at Figure \ref{fig:ess-triple} for $Q^2=100$ GeV$^2$ and $Q^2=10000$ GeV$^2$. In addition, the predictions for the triple-pole residues for $F_2^p$ are shown at figure \ref{fig:formfact} together with the $\chi^2$ per point of the fit.

We clearly see that the parametrisation works very well for valence quarks at all values of $Q^2$, as well as for the $u_s$ and $d_s$ distributions. For heavy quarks and gluons, although the $\chi^2/nop$ remains less than 1, there are some discrepancies between the DGLAP essential singularity and the triple-pole fit at small $x$ and large $Q^2$. However, these differences are only present for $\sqrt{s}> 3$ TeV, which means that it is impossible to distinguish between the two approaches in present experimental measurements. This high-energy region should be reached at the LHC and should provide very useful information to distinguish between the different models.

Moreover, we have also tried to estimate the uncertainties on the parton densities in another way. We have considered the uncertainties obtained from the DGLAP fit at $Q^2=Q_0^2$:
\[
q(x,Q_0^2)-\Delta q(x,Q_0^2) \le q(x,Q_0^2) \le q(x,Q_0^2)+\Delta q(x,Q_0^2).
\]
We can then evolve $q(x,Q_0^2)\pm\Delta q(x,Q_0^2)$ and using these quantities to obtain the uncertainties at all values of $Q^2$. If we do so, we obtain very similar conclusions.

\section{Conclusions and perspectives}

We have seen that we can use Regge theory to constrain the initial parton densities at $Q^2=Q_0^2$ and obtain the distributions at higher virtualities with the DGLAP evolution equation. In this approach, Regge theory is used to describe the low-$Q^2$ data and QCD applies at large $Q^2$. In such a way, the complex-$j$-plane singularities are common to parton distribution functions in the initial condition and to soft amplitudes which provides a unified description at high-energy in the soft region.

We have also shown in this paper that it is possible to  {\em define} the parton distributions in the low-$Q^2$ region and to parametrise them using Regge theory. This parametrisation is useful to describe the DIS structure functions but should be used with care. Actually, since factorisation is not proven at small $Q^2$, we cannot ensure that the parton distributions can be extended to $Q^2=0$. Using our parametrisation to describe processes such as jet production may be incorrect. 

Considering the low-$Q^2$ parametrisation together with the Global QCD fit, we have a combined description of the hadronic structure functions over the whole $Q^2$ range. This model, consistent with DGLAP evolution and with Regge theory, reproduces the experimental measurements with a very good $\chi^2$.

In addition, we extended the approach of \cite{Soyez:2002nm} to $x=1$ using only forward evolution. We have not applied the techniques developed in \cite{Soyez:2003sr} and extracted the $Q^2$ behaviour of the fitted parameters by combining forward and backward evolution. The reason is that, even with a few parameters, there often exist multiple minima and it is quite hard to obtain a continuous result for all parameters. This situation is expected to be even worse with the parametrisation used here due to the larger number of parameters. Hence, in order to test the compatibility between the DGLAP-evolved parton distributions and a triple-pole parametrisation, we have shown that the parton densities can be approximated by a $\log^2(1/x)$ behaviour at small $x$ and large $Q^2$. This approximation works very well up to $\sqrt{s}\approx 3$ TeV, and at which point it deviates from DGLAP for the heavy quarks and the gluons. This means that we expect high-energy corrections to be important in this domain and that the LHC should provide very useful information to distinguish between the different high-energy models. In this high-energy region, one should expect contributions from the BFKL equation as well as unitarity and saturation effects.

Finally, a NLO analysis will be performed in the near future. This gives a much more reliable description of the data, allows a more complete comparison with other parametrisations and gives a description of the $F_c$ and $F_L$ structure functions.

\vspace{1.5cm}
\begin{center}
{\bf Acknowledgements}
\end{center}
First of all, I would like to thank J.R. Cudell for very useful discussions and suggestions. I am also very grateful to L. Favart and J. Stirling. Finally, I would like to thank Y.K. Yang for fruitful discussions concerning the CCFR measurements. This work is supported by the National Fund for Scientific Research (FNRS), Belgium.

\begin{landscape}
\begin{table}
\begin{center}
{\footnotesize{
\begin{tabular}{|c|c|cccccccccccccc|}
\hline
  Param. & global & $A$ & $B$ & $C$ & $D_u$ & $D_d$ & $D_s$ & $A_g$ & $B_g$ & $\gamma_u$ & $\gamma_d$ & $b_u$ & $b_d$ & $b$ & $\eta$ \\
\hline
$A$        &0.99650& 1.000&-0.895&-0.015& 0.733& 0.631& 0.007&-0.363& 0.268& 0.365&  0.304& 0.018&-0.022& 0.442&-0.368\\
$B$        &0.99824&-0.895& 1.000& 0.025&-0.864&-0.668&-0.006& 0.229&-0.148&-0.567& -0.494&-0.081&-0.042&-0.589& 0.562\\
$C$        &0.28980&-0.015& 0.025& 1.000& 0.005& 0.005& 0.000& 0.001&-0.002& 0.001&  0.001& 0.001& 0.000& 0.000& 0.002\\
$D_u$      &0.99222& 0.733&-0.864& 0.005& 1.000& 0.822& 0.006& 0.086&-0.177& 0.524&  0.521& 0.267& 0.085& 0.876&-0.460\\
$D_d$      &0.98233& 0.631&-0.668& 0.005& 0.822& 1.000& 0.005&-0.071&-0.009& 0.110&  0.040& 0.397&-0.243& 0.762&-0.009\\
$D_s$      &0.02647& 0.007&-0.006& 0.000& 0.006& 0.005& 1.000& 0.001&-0.002& 0.005&  0.005& 0.004& 0.000& 0.008&-0.004\\
$A_g$      &0.99616&-0.363& 0.229& 0.001& 0.086&-0.071& 0.001& 1.000&-0.980& 0.346&  0.380& 0.253& 0.199& 0.356&-0.291\\
$B_g$      &0.99649& 0.268&-0.148&-0.002&-0.177&-0.009&-0.002&-0.980& 1.000&-0.403& -0.443&-0.317&-0.228&-0.464& 0.332\\
$\gamma_u$ &0.99956& 0.365&-0.567& 0.001& 0.524& 0.110& 0.005& 0.346&-0.403& 1.000&  0.900& 0.253& 0.318& 0.518&-0.980\\
$\gamma_d$ &0.99665& 0.304&-0.494& 0.001& 0.521& 0.040& 0.005& 0.380&-0.443& 0.900&  1.000& 0.081& 0.648& 0.552&-0.893\\
$b_u$      &0.97469& 0.018&-0.081& 0.000& 0.267& 0.397& 0.004& 0.253&-0.317& 0.253&  0.081& 1.000&-0.232& 0.537&-0.073\\
$b_d$      &0.97329&-0.022&-0.042& 0.000& 0.085&-0.243& 0.000& 0.199&-0.228& 0.318&  0.648&-0.232& 1.000& 0.174&-0.332\\
$b$        &0.99685& 0.442&-0.589& 0.002& 0.876& 0.762& 0.008& 0.356&-0.464& 0.518&  0.552& 0.537& 0.174& 1.000&-0.389\\
$\eta$     &0.99965&-0.368& 0.562&-0.001&-0.460&-0.009&-0.004&-0.291& 0.332&-0.980& -0.893&-0.073&-0.332&-0.389& 1.000\\
\hline
\end{tabular}}}
\end{center}
\caption{Fitted parameters correlation coefficients.}\label{tab:xto1-corel}
\end{table}
\end{landscape}

\begin{appendix}
\section{Momentum sum rule and gluon distribution}

In this appendix, we shall give the expression of the constant in the gluon distribution, constrained by the momentum sum rule. Recall that we have, at $Q^2=Q_0^2$,
\begin{eqnarray}\label{eq:initq}
xu_V(x)       & = & \frac{2}{N_u}x^\eta(1+\gamma_u x)(1-x)^{b_u},\nonumber\\
xd_V(x)       & = & \frac{1}{N_d}x^\eta(1+\gamma_d x)(1-x)^{b_d},\nonumber\\[-3mm]
& & \\[-3mm]
x\bar{q}_i(x) & = & \left[A\log^2(1/x)+B\log(1/x)+C+D_ix^\eta\right](1-x)^b,\nonumber\\
xg(x)         & = & \left[A_g\log^2(1/x)+B_g\log(1/x)+C_g\right](1-x)^{b+1}\nonumber,
\end{eqnarray}
where $N_q$ is given by equation \eqref{eq:qnumc}. We shall use momentum conservation to constrain the constant term $C_g$ in the gluon distribution. Let us first introduce the special functions that we need. The Euler Gamma function is defined by
\[
\Gamma(x) = \int_0^\infty dt\,t^{x-1} e^{-t}.
\]
We can then introduce the Beta function $B(x,y)$, the digamma function $\Psi(x)$ and the polygamma function $\Psi^{(m)}(x)$ related to the gamma functions by the following formul\ae
\begin{eqnarray*}
  B(x,y)        & = & \frac{\Gamma(x)\Gamma(y)}{\Gamma(x+y)},\\
  \Psi(x)       & = & \frac{\partial_x \Gamma(x)}{\Gamma(x)}, \\
  \Psi^{(m)}(x) & = & \partial_x^m\Psi(x).
\end{eqnarray*}
With these definitions, the momenta carried by the distributions \eqref{eq:initq} are given by the following expressions:
\begin{eqnarray*}
p_{u_V} & = & \frac{2\eta}{b_u+\eta+1}\left(1+\gamma_u\frac{\eta+1}{b_u+\eta+2}\right)\left(1+\gamma_u\frac{\eta}{b_u+\eta+1}\right)^{-1},\\
p_{d_V} & = & \frac{\eta}{b_d+\eta+1}\left(1+\gamma_d\frac{\eta+1}{b_d+\eta+2}\right)\left(1+\gamma_d\frac{\eta}{b_d+\eta+1}\right)^{-1},\\
p_{\bar{q}_i} & = & \frac{1}{b+1}\left(A\left\{\left[\gamma_E+\Psi(b+2)\right]^2-\Psi^{(1)}(b+2)+\frac{\pi^2}{6}\right\}+B\left[\gamma_E+\Psi(b+2)\right]+C\right)\\
              & + & D_i B(b+1,\eta+1),\\
p_{g} & = & \frac{1}{b_g+1}\left(A_g\left\{\left[\gamma_E+\Psi(b_g+2)\right]^2-\Psi^{(1)}(b_g+2)+\frac{\pi^2}{6}\right\}+B_g\left[\gamma_E+\Psi(b_g+2)\right]+C_g\right).
\end{eqnarray*}
From the proton, momentum conservation gives
\[
p_g + p_{u_V} + p_{d_V} + 2\left(p_{\bar{u}}+p_{\bar{d}}+p_{\bar{s}} \right) = 1,
\]
and we finally obtain
\begin{eqnarray*}
c_G & = & A_g\left\{\left[\gamma_E+\Psi(b_g+2)\right]^2-\Psi^{(1)}(b_g+2)+\frac{\pi^2}{6}\right\}+B_g\left[\gamma_E+\Psi(b_g+2)\right] \\
    & + & (b_g+1)\left[ 1-p_{u_V} - dp_{_V} - 2\left(p_{\bar{u}}+p_{\bar{d}}+p_{\bar{s}} \right)\right].
\end{eqnarray*}

\end{appendix}

\end{document}